\documentclass[useAMS,usenatbib, usegraphix]{mn2e}
\usepackage{graphicx}
\usepackage{lscape}
\usepackage{rotating}
\usepackage{aasmacros, amsmath}
\usepackage{placeins}
\usepackage{times} 
\usepackage[T1]{fontenc}

\newcommand{\mdot}{M_{\odot}}

\title[Galaxy and Mass Assembly (GAMA): Projected Galaxy Clustering]{Galaxy and Mass Assembly (GAMA): Projected Galaxy Clustering}
\author[D. J. Farrow ]{D. J. Farrow$^{1,2}$\thanks{E-mail:dfarrow@mpe.mpg.de}, Shaun Cole$^{1}$, Peder Norberg$^{1,3}$, 
N. Metcalfe$^{3}$,  I. Baldry$^{4}$, \and  Joss Bland-Hawthorn$^{5}$, Michael J. I. Brown$^{6}$, 
A. M. Hopkins$^{7}$ Cedric G. Lacey$^{1}$, \and J. Liske$^{8}$, Jon Loveday$^{9}$, David P. Palamara$^{6}$, 
A.S.G. Robotham$^{10}$, Srivatsan Sridhar$^{9,11}$  \\
\\
$^{1}$ICC, Department of Physics, University of Durham, South Road, DH1 3LE, U.K.\\
$^{2}$Max-Planck-Institut f{\"u}r extraterrestrische Physik, Postfach 1312 Giessenbachstrasse, 85741 Garching, Germany\\
$^{3}$CEA, Department of Physics, University of Durham, South Road, DH1 3LE, U.K.\\
$^{4}$Astrophysics Research Institute, Liverpool John Moores University, IC2, Liverpool Science Park, 146 Brownlow Hill, Liverpool, L3 5RF\\
$^{5}$Sydney Institute for Astronomy, School of Physics A28, University of Sydney, NSW 2006, Australia \\
$^{6}$School of Physics and Astronomy, Monash University, Clayton, Victoria 3800, Australia\\
$^{7}$Australian Astronomical Observatory, PO Box 915, North Ryde, NSW 1670, Australia \\
$^{8}$Hamburger Sternwarte, Universit{\"a}t Hamburg, Gojenbergsweg 112, 21029 Hamburg, Germany \\
$^{9}$Astronomy Centre, University of Sussex, Falmer, Brighton BN1 9QH \\
$^{10}$ICRAR M468, University of Western Australia, 35 Stirling Hwy, Crawley WA 6009 Australia\\ 
$^{11}$Laboratoire J. L. Lagrange, Observatoire de la C\^{o}te d'Azur, BP4429, 06300, Nice, France (present address)\\
}

\begin{document}
\date{}
\pagerange{\pageref{firstpage}--\pageref{lastpage}} \pubyear{2015}

\maketitle
\label{firstpage}

\begin{abstract}
We measure the projected 2-point correlation function of galaxies in the 180 deg$^{2}$ equatorial regions of the GAMA II survey, for four different redshift slices between 
$z=0.0$ and $z=0.5$. To do this we further develop the \citet{cole11} method of producing suitable random catalogues for the calculation of
correlation functions. We find that more $r$-band luminous, more massive and redder galaxies are more clustered. We also find that red galaxies 
have stronger clustering on scales less than $\sim 3\,h^{-1}$Mpc. We compare to two different versions of the {\sc galform} galaxy formation 
model, Lacey et al (in prep.) and \citet{gonzalez14}, and find that the models reproduce the trend of stronger clustering for more massive galaxies. However, the 
models under predict the clustering of blue galaxies, can incorrectly predict the correlation function on small scales and
under predict the clustering in our sample of galaxies with $\sim 3 L_{r}^*$. We suggest possible avenues to explore to improve these clustering predictions. 
The measurements presented in this paper can be used to test other galaxy formation models, and we make the measurements available online to facilitate this.  
\end{abstract}

\begin{keywords}
large-scale structure of Universe, galaxies: evolution, galaxies: formation
\end{keywords}

\section{Introduction}
The two-point auto-correlation function is a widely used statistical description of the spatial distribution of galaxies. In the standard
model of cosmology, the shape of this function is set both by the mass of dark matter haloes in which a particular galaxy sample resides, and baryonic processes which can 
change the spatial distribution of galaxies on smaller scales. Its dependence on galaxy properties is well established, at low redshifts, by 
large area spectroscopic surveys such as SDSS \citep{york00,strauss02} and 2 Degree Field Galaxy Redshift Survey \citep[2dFGRS;][]{colless01}. The amplitude of the auto-correlation function of galaxies is seen to be
strongly dependent on luminosity, stellar mass and colour. At low redshifts, brighter, redder and more massive galaxies have been observed to be more strongly 
clustered \citep[e.g.][]{norberg01, norberg02, li06, zehavi02, zehavi05, zehavi11, christodoulou12}.

In the higher redshift Universe, small ($<1$~deg$^2$) but deep spectroscopic surveys have also measured the clustering of galaxies. The DEEP2 survey has 
been used to demonstrate that the colour dependence of galaxy clustering is already in place at $z\sim1$, whilst within a red or blue sample of galaxies, 
clustering is insensitive to luminosity over the range $20.2<M_{B}<21.8$ \citep{coil08}. Compared with lower redshift SDSS data, the DEEP2 measurements of the 
clustering of brighter and more massive galaxies has a larger amplitude than expected from scaling the low redshift measurements using linear theory \citep{coil08, li12}. 
This can be interpreted as evidence of significant bias evolution for these galaxies \citep{coil08, li12}.  Another example of a small area, deep 
spectroscopic survey is the VIMOS-VLT Deep Survey (VVDS), which found a sharp increase in the amplitude of galaxy clustering around the 
characteristic magnitude of the sample's luminosity function \citep{pollo06}. VVDS also found, in agreement with DEEP2, that the 
clustering of massive galaxies, with stellar mass $>10^{10.5}h^{-2}\mathrm{\mdot}$, at $z\sim1$ can only be reconciled with lower redshift measurements if their bias evolves 
significantly \citep{meneux08}. Results from another small area, deep survey, zCOSMOS, show that this is true for galaxies more 
massive than $>10^{10}h^{-2}\mathrm{\mdot}$ \citep{meneux09}. The VIPERS survey has also found that more luminous and more massive galaxies are
more clustered at redshifts of $0.5<z<1.1$ than their fainter, less massive counterparts \citep{torre13, marulli13}. More recently, the 9 deg$^{2}$ 
PRIMUS survey \citep{coil11, cool13} has found red galaxies are more clustered than blue galaxies, and the relationship between clustering amplitude and sample 
luminosity is in place in the redshift slices: $0.2<z<0.5$ and $0.5<z<1.0$ \citep{skibba14}. 

In addition to spectroscopic surveys, surveys relying on photometric redshifts have also explored the clustering of galaxies. Galaxies with red colours have
been shown to have steeper correlation functions than bluer galaxies at $z \sim 0.5$ in the 0.78~deg$^{2}$ COMBO-17 survey \citep{phleps06}, and the 1.5~deg$^{2}$ UltraVISTA
survey demonstrated that more massive galaxies are more clustered between $0.5<z<2.5$ \citep{mccracken14}. Photometric surveys have also studied the evolution of galaxy clustering.
In the 6.96 deg$^{2}$ Bo{\"o}tes field \citet{brown08} found, at $0.2<z<1.0$, that red galaxies fainter than $M_{B} = 20 - 5\log_{10}h$ showed 
no luminosity dependence of the amplitude of their clustering, whilst the converse is true for red galaxies brighter than this. Studies in the Bo{\"o}tes field also demonstrated 
luminous red galaxies display little evolution in the amplitude of their clustering between $z=0.5$ and $z=0.9$ \citep{white07, brown08}. This observation, which suggests the clustering of these galaxies 
evolves slower than the underlying dark matter distribution, can be explained by the removal of highly-biased satellite galaxies by merging or disruption \citep{white07, brown08}. Whilst bright, red galaxies display little evolution in clustering 
strength with redshift, fainter galaxies ($-22<M_{B} - 5\log_{h}<-19$) have been shown to have a decreasing clustering amplitude between $z=0.4$ and $z=1.2$ in the CFHTLS photometric redshift survey \citep{mccracken08}. 

The photometric surveys at high redshift are still fairly small ($<10$~deg.$^2$). A possible exception to the trade-off between large area and high redshift is the work of \citet{guo14}; here the authors used the CMASS sample of SDSS to
show that clustering was stronger for brighter and redder galaxies at $z \sim 0.5$. However, the complicated selection of the CMASS sample limits their work
to a very narrow luminosity and redshift range.   

The GAMA spectroscopic survey offers a new window onto the clustering of galaxies and its evolution with redshift. It has a larger area (the equatorial regions we use total $180$ deg$^{2}$) 
than the deep surveys but has spectra of much fainter galaxies (2 magnitudes) than large-area surveys like SDSS. As such it complements both types of survey, by enabling 
detailed clustering measurements at an intermediate epoch. Whilst our redshift range overlaps with the PRIMUS low-redshift sample, our larger area allows us to split 
the data into finer redshift bins and consider brighter, rarer objects. In this paper we study the projected two-point correlation function (2PCF) of galaxies as a 
function of their luminosity, stellar mass and colour, in four different redshift bins, in order to fully appreciate any redshift 
evolution. To aid in the physical interpretation of our results we compare them to different versions of the  semi-analytic galaxy formation model 
{\sc galform}, specifically \citet{gonzalez14} (hereafter G14 model) and Lacey et al (in prep.) (hereafter L14 model). We quantitatively compare 
the relative merits of the different versions of the {\sc galform} model and discuss the impact of this on our understanding of galaxy formation. To see
a complementary approach, we refer the reader to Palamara et al. (in prep.), who fit halo occupation distribution models \citep[HOD models; see e.g.][]{zheng05} to GAMA.

This paper is organised as follows. Section~\ref{sec:dataGAMA} introduces the GAMA data and the galaxy formation model we use, along with details of how 
we calculate luminosity, mass and rest-frame colour. Section \ref{sec:methodGAMA} presents our method of generating a random catalogue and measuring 
clustering. In Section \ref{sec:resultsGAMA} we present our results, before discussing them and concluding in Section \ref{sec:concGAMA}. We present 
our results using units of a fiducial $\Lambda$CDM cosmology with $\Omega_{\rm m}=0.25$, $\Omega_{\Lambda}=0.75$ and $H=100h$~km~s$^{-1}$~Mpc$^{-1}$.

\section{Data and Models}\label{sec:dataGAMA}
\subsection{The Galaxy and Mass Assembly survey}
The GAMA survey is a spectroscopic and multi-wavelength survey of galaxies carried out on the Anglo-Australian telescope \citep{driver11, liske15}. In this work we 
utilise the main $r$-band limited data from the GAMA II equatorial regions, which consists of a highly complete ($>98$\%) spectral catalogue of galaxies selected from the SDSS DR7 \citep{abazajian09} to 
have $r_{\rm{petro}}<19.8$. The older GAMA I survey had a shallower limit of $r<19.4$ in two of the regions. Target fields are repeatedly observed in a way that removes biases against close pairs \citep{robotham10}, 
avoiding such biases is ideal for clustering measurements.  Star/galaxy separation is based on the difference between an object's model and PSF magnitudes in SDSS DR7 data and, where UKIDSS photometry is available, the object's optical 
and infrared colours. Further details of the GAMA survey are given in \citet{baldry2010}, \citet{robotham10}, \citet{driver11} and \citet{liske15}. The data are split over three $12 \times 5$~deg$^{2}$ fields 
centered at $9^{h}$ (G09), $12^{h}$ (G12) and $14.5^{h}$ (G15) R.A. and approximately $\delta=0$~degrees declination. 

\subsubsection{Redshifts}
Redshifts for GAMA objects were measured automatically, using the software {\sc autoz} as described in \citet{baldry14}. \citet{liske15} find $0.2\%$ of the sample are 
expected to have an incorrect redshift. The median velocity uncertainty of the measured redshifts is 27 $\rm{km\,s^{-1}}$ 
\citep{liske15}. The redshifts are taken from a table called {\sc DistancesFramesv12} in a GAMA Data Management Unit (DMU). These redshifts have been corrected for 
the local flow using the model of \citet{tonry00}, smoothly tapered to the CMB restframe for $z \ge 0.03$. 
 
\subsubsection{Quality cuts}
In addition to the redshift quality cut, $nQ \geq 3$, we also only consider galaxies in regions with completeness greater than 80\% using the GAMA angular 
completeness mask \citep{driver11, liske15}. We additionally only select objects with {\sc vis\_class$=0$}, {\sc vis\_class$=1$} or { \sc vis\_class$=255$}, which 
removes objects which upon visual inspection do not show any evidence of galaxy light or appear to be part of another galaxy \citep[]{baldry2010}.

\subsubsection{Magnitudes}
GAMA combines data from a large number of ground and space based telescopes and so has a very wide wavelength range, from X-ray to radio. In this work 
we use optical photometry from SDSS DR7 imaging data. To define luminosity samples we use SDSS Petrosian magnitudes \citep{petrosian1976}, as the GAMA 
selection used Petrosian magnitudes. To define colours, and when estimating stellar mass, we use SDSS model magnitudes as these are often more suitable 
for colour terms (see the SDSS DR7 photometry webpage\footnote[1]{http://classic.sdss.org/dr7/algorithms/photometry.html, accessed 19/8/14}). Following SDSS 
conventions, we will label model magnitudes using the letter associated with the bandpass in which the magnitude was measured. We took these magnitudes 
from the table called {\sc TilingCatv42} in a GAMA DMU, where they are replicated. 

GAMA also has a set of magnitudes measured in apertures with a matched size across the different bands \citep{hill11}, these magnitudes have been shown 
to be superior to model magnitudes when calculating a galaxy's stellar mass using spectral energy distribution (SED) fitting \citep{taylor11}. Our adoption of model magnitudes 
was in order to avoid the small number of failures whose spatial positions have not been mapped in the \citet{hill11} catalogue. Model magnitudes have been widely used to 
define colours in SDSS publications, and should be sufficient to separate red and blue galaxies in this work.

\subsubsection{k-corrections and evolution corrections}\label{sec:kcorr}
To compute Petrosian absolute $r$-band magnitudes, $M_{r} - 5\log_{10} h$ (hereafter $M_{r,h}$), from the observed apparent Petrosian $r$-band magnitudes, $m_{r}$, we apply both a k-correction,
$k_{r}$, and a luminosity evolution correction. The latter correction is necessary in order to compare a similar population of galaxies across time, as the 
luminosity of galaxies evolves. The relation we use is
\begin{equation}
\label{eq:mtoM}
M_{r,h} = m_{r} - k_{r}(z) + Q(z - z_{\mathrm{ref}}) - 5\log_{10}(D_{L}(z)) - 25 ,
\end{equation}    
where $Q$ is the luminosity evolution parameter, $D_{L}$ the luminosity distance in $h^{-1}$~Mpc and $z_{\rm{ref}}$ is a reference redshift, for 
which we adopt $z_{\rm{ref}}=0.0$. The parameterisation of luminosity evolution we use is commonly adopted in the literature \citep[e.g.][]{lin99, loveday12}.

At this stage we will also introduce a parameterisation of density evolution common to luminosity function studies. The $P$ parameter 
\citep[e.g.][ and references therein]{lin99, loveday12} parameterises the density evolution of a population of galaxies via
\begin{equation}
\phi^{*}(z) = \phi^{*}(z=0)10^{0.4Pz}
\end{equation} 
where $\phi^{*}(z)$ is the characteristic number density of galaxies at redshift $z$. \citet{loveday12} have 
fit evolving luminosity functions to  a shallower version of the GAMA data, with a limit of $r_{\rm{petro}}<19.4$ in G09 and G15. 
They find that $Q$ and $P$ are very degenerate \citep[see also][]{farrowThesis}. In \citet{loveday15} adopting a 
value of $Q=1.45$ resulted in very little need for density evolution \citep[see also][]{farrowThesis}. We therefore adopt $Q=1.45$, which 
combined with our method of generating a random catalogue, removes the need to correct for density evolution (see Section \ref{sec:randGAMA}).
Note that in \citet{loveday15} a different set of parameters are favoured. However in Appendix~\ref{sec:cTest} we show that adopting the parameters $P=1.45$ and $Q=0.81$, 
which we took from an earlier draft of \citet{loveday15}, affects the correlation function in a way far smaller than the errors.

The k-corrections are derived from the GAMA DMU {\sc kcorr\_z00v04}, which was produced using the method set out in \citet{loveday12}. The 
\citet{loveday12} k-corrections are found by using the code {\sc kcorrect\_v4\_2} \citep{blanton07} to fit each galaxy's $u$, $g$, $r$, $i$, 
and $z$-band SDSS model magnitudes with SED templates. For many applications the maximum redshift at which an object fulfils the selection criteria 
of the survey, $z_{\rm{max}}$, is needed. For this the k-correction as a function of redshift is required for each galaxy. To enable fast computation, 
\citet{loveday12} fit a 4th order polynomial to the k-correction of each galaxy as a function of redshift. The rms difference between the {\sc kcorrect} 
estimates of k-correction and the polynomial fits to them is less than 0.01 magnitudes for all bands \citep{loveday12}. We further speed up the 
k-correction process by using average polynomials from \citet{tamsyn14}, who computed the median of the \citet{loveday12} k-correction polynomials 
for galaxies in seven, $(g-r)$ rest frame colour bins, hereafter labelled $(g-r)_0$. The benefit of this is a reduction in noise introduced by fitting 
each galaxy with an individual model \citep{tamsyn14}. Rest-frame colours are still estimated from the individual k-corrections of \citet{loveday12}.

\subsubsection{Stellar mass}
As well as luminosity and colour, we also want to use GAMA II to measure clustering as a function of stellar mass. We use the relation between colour 
and stellar mass found for GAMA I data by \citet{taylor11}, namely
\begin{equation}
\label{eq:mass}
\mathrm{log}_{10}M_{*} (h^{-2}\mathrm{\mdot}) = 1.15 + 0.7(g - i)_{0} - 0.4(M_{i} - 5\mathrm{log_{10}h})
\end{equation}
where $M_{i}$ is the rest frame $i$-band absolute model magnitude and $(g-i)_{0}$ is the rest frame colour. This relation was found by \citet{taylor11} 
from individual estimates of the stellar mass of GAMA I galaxies, produced by fitting stellar population synthesis models to the optical GAMA data 
\citep[details in][]{taylor11}. These mass estimates should have a statistical 1$\sigma$ accuracy of around 0.1~dex \citep{taylor11}. The 
original relation used the GAMA matched-aperture magnitudes for the colours, whilst we use SDSS model magnitudes. We have tested that this does not add significant biases or 
scatter to our computed stellar masses. The resultant offset between our mass-to-light ratios and the ones utilizing the GAMA matched aperture colours with 
Eq.~\ref{eq:mass} is only 0.01~dex with an interquartile range of 0.1~dex and 10\% of outliers with a difference of $>0.14$~dex. Note that the problem
\citet{taylor11} discovered with masses using model magnitudes arose during full SED fitting, not using this colour relation, and were reported to
be driven by the $u$-band, which has no influence on Eq.~\ref{eq:mass}. 

\subsection{Galaxy Formation Models \& Lightcones}\label{sec:mock}
\subsubsection{The {\sc galform} model}
We compare our observations to a semi-analytic galaxy formation model called {\sc galform}. The {\sc galform} model was first presented in 
\citet{cole2000}, but it grew from a number of previous attempts to model galaxy formation. These models all assume galaxies form in dark matter haloes,
 and then use analytical prescriptions to approximate key galaxy formation processes \citep[e.g.][]{white78, white91, kauffmann93,cole94}. The first 
{\sc galform} model has prescriptions for the cooling rate of gas in dark matter halos, a star formation rate based on available cold gas content, the 
attenuation of star formation by supernovae feedback and the merging of satellite galaxies to the main galaxy \citep{cole2000}. Many of these 
prescriptions have free parameters, controlling factors like the strength of supernovae feedback, that account for complicated physics not fully 
understood or analytically described. These parameters are tuned such that the model fits a set of observational constraints at low redshift. 
None of the models used here were tuned to reproduce clustering measurements.

The models require the history of when a dark matter halo formed and merged, sometimes called a halo's `merger tree' \citep{lacey93}. For this purpose 
the models we present here use an N-body simulation based on the Millennium Simulation \citep{springel05}, but run in the more up-to-date WMAP7 
cosmology \citep[][]{komatsu11}. 

\subsubsection{Model Lightcones}\label{sec:lightcones}
The positions and velocities of particles in the simulation are output at epochs spaced in expansion factor, hereafter ``snapshots''.  Mock catalogues of 
galaxies from {\sc galform} runs on the N-body simulation are produced using the code of \citet[][]{merson13}. This works by interpolating between the snapshots to find the positions of 
galaxies when they enter the simulated observer's lightcone, i.e. when the light from the galaxy reaches the observer.

From the simulations 26 realisations of each model were created by changing the position and orientation of the virtual observer\footnote[2]{These mock catalogues will be made available
from the VirgoDB: http://icc.dur.ac.uk/data/}. These multiple realisations allow us to improve our understanding of sample variance (Section~\ref{sec:jackknife}), and minimise its effect 
on predictions by plotting mean measurements from the mock realisations (Section~\ref{sec:resultsGAMA}). One downside is the simulation used has a limited volume, 
$12.5 \times 10^{7}~(\mathrm{Mpc}/h)^3$, as such the 
upper-intermediate and highest redshift slices will over-sample, over 26 realisations, the simulation by at least a factor of 3 and 8. Note this is the total oversampling when you consider  
a sum over all the realisations, an individual realisation is much smaller than the simulation box. Also mitigating this effect is the fact that the 
location and orientation of the observer is randomly assigned, in this way repeated structures will be observed at different redshifts and orientations, and be sampled by different galaxies. 
In the lower-intermediate redshift slice the volume of the 26 realisations is only 4\% larger than the simulation, which should result in very little oversampling. How the oversampling 
affects results will be discussed in later sections.

The next sections describe the specifics of the two versions of {\sc galform} we consider. A summary of the relevant {\sc galform} models can be seen in Table~\ref{table:models}.
\begin{table*}
\caption{Summary of the different {\sc galform} models mentioned in this paper.}
\begin{tabular}{l l l l l l}
\hline\hline
Model & Parent Model & Cosmology & Key Features\\
\hline
\citet{baugh05} & \citet{cole2000} & WMAP1 & Different IMFs in starbursting galaxies.\\
Bower et al (2006) & Cole et al (2000) & WMAP1 & AGN feedback in hydrostatic haloes.\\
Lagos et al (2012) & \citet{bower06} & WMAP1 & New star formation law accounting for  \\&&&  fraction of hydrogen in molecular state.\\
G14: Gonzalez-Perez et al (2014) & Lagos (2012) & WMAP7 & Updated cosmology.  \\
L14: Lacey et al (in prep.) & Baugh et al (2005) \& Lagos et al (2012) &  WMAP7 & Combines varying IMF  \\&&& and new star formation law. \\
\hline
\end{tabular}\label{table:models}
\end{table*}
\subsubsection{The G14 model}
The \citet{gonzalez14} (G14) model has the same physical prescriptions as the \citet{lagos12} model, which is itself and extension of the \citet{bower06} 
model. The \citet{bower06} model added AGN feedback to the \citet{cole2000} model in order to better reproduce galaxy colours, and to decrease the 
number of galaxies predicted to lie in the bright end of the luminosity function. The \citet{lagos12} model made the star formation rate in the disk of the
galaxy proportional to the fraction of the gas that was molecular, by using the empirical star formation law of \citet{blitz06} \citep[see][for details]{lagos11, lagos12}.

The update of the G14 model over \citet{lagos12} is that it uses merger trees from WMAP7 cosmology. WMAP7 cosmology has
$\sigma_{8}=0.81$, this is smaller than in the WMAP1 cosmology of \citet{lagos12} ($\sigma_{8}=0.9$), so free parameters in the physical prescriptions had 
to be retuned \citep{gonzalez14}. Tunable parameters were adjusted in order for the predictions to still match the rest-frame $b_{J}$ and K-band luminosity functions at $z=0$, and give reasonable evolution 
of the UV and K-band luminosity functions \citep{gonzalez14}.  

The G14 model converts from star formation history and the IMF to a spectral energy distribution (SED) using an updated version of the stellar
population synthesis (SPS) model from \citet{bruzual93}.

\subsubsection{The L14 model}
The Lacey et al (in prep.) (L14) model also combines the improvements to star formation rate calculations made in the \citet{lagos12} model and the AGN feedback model of 
\citet{bower06}. It was also run on the new WMAP7 N-body simulation. Another difference of the L14 model is its initial mass function (IMF). The G14 model uses a 
\citet{kennicutt83} IMF for all stars formed. The L14 model instead adopts a top-heavy IMF in starbursting galaxies, as developed in, but less extreme than, the \citet{baugh05} model. 
This change was motivated to bring the predictions of sub-mm galaxy number counts into closer agreement with observations \citep{baugh05}. 

A further difference in the L14 model which could have an impact on clustering predictions is the treatment of merging 
satellite galaxies. The models we present here all use an analytic approximation, from \citet{cole2000}, to calculate how long it takes for an accreted 
satellite to merge with the central galaxy. This approximation assumes satellites enter the halo on orbits randomly selected from the distribution of 
satellite orbits given in \citet{tormen97}, before the orbit decays due to dynamical friction. The \citet{cole2000} method calculates the time for the 
orbit to decay using the formula for Chandrasekhar dynamical friction in an isothermal sphere given in \citet{lacey93}. The L14 model, on the other 
hand, uses the \citet{jiang08, jiang14} formula for the timescale. The formula has been empirically modified using N-body simulations in order to account 
for the tidal stripping of the accreting haloes. Note that \citet{campbell15}, motivated by the work of \citet{contreras13}, additionally develops the
 model to track the positions of accreted subhaloes in the simulation, rather than use an analytic approximation. We do not attempt modifications to the 
models in this paper as our focus is on presenting and interpreting the observations.  

A final difference between this model and G14 is the adoption of the \citet{maraston05} SPS model, which differs from the \citet{bruzual93} model in
its treatment of thermally-pulsating asymptotic giant branch stars (TP-AGB). These stars are important in the near-IR. The near-IR corresponds to the $r$ and $i$-bands we 
use for luminosity and colour measurements \citep{maraston05}.

Note that the version of the L14 model we use is slightly different to more recent versions of the Lacey et al (in prep.) model. The differences are tiny, however,
and should not have any impact at the redshifts we consider. 

\subsubsection{Model Stellar Masses}
\begin{figure}
\includegraphics[width=0.45\textwidth]{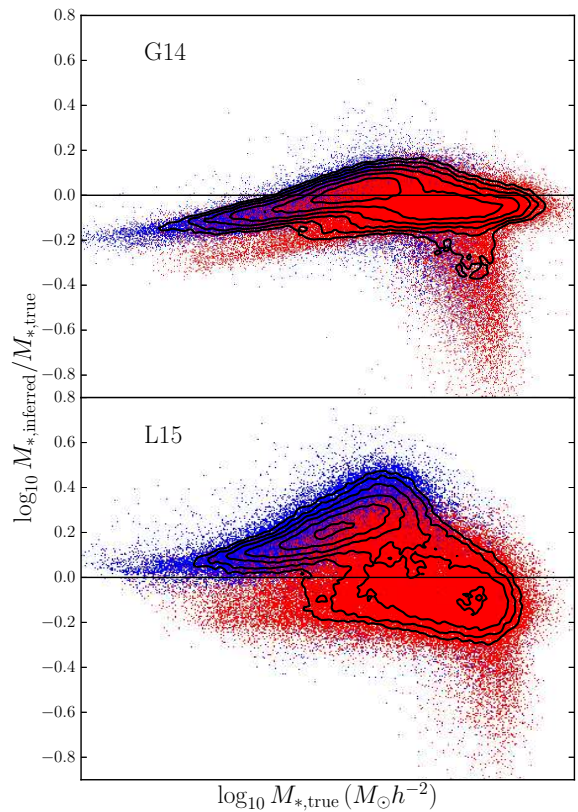}
\caption{The inferred mass, from the Taylor et al (2011) colour to mass relation, versus the true mass of mock galaxies, for the G14 (top) and 
L14 (bottom) models. The red and blue points represent red and blue galaxies, as selected by our colour cuts (Sec.~\ref{sec:samp}). The contours are spaced evenly in log-space,
between 5$\times 10^4$ and 50$\times 10^4$ galaxies dex$^{-2}$. Inferring the mass using observational relations, rather than using the true model mass can make a significant 
difference and is in this context model dependent.}\label{fig:massCmp}
\end{figure}

We take stellar masses from the model, but rescale them by $h=0.702$ (the model cosmology) in order to convert the physical units of the model, $\mathrm{M_{\odot}} h^{-1}$, to
the units of the observations, $\mathrm{M_{\odot}} h^{-2}$. \citet{mitchell13} demonstrates that applying broad-band SED fitting to model galaxy photometry can 
give a biased estimate of the true model galaxy's mass. In order to investigate what effect this may have we show the inferred mass, from Eq~\ref{eq:mass}, against the
true model mass in Fig.~\ref{fig:massCmp}. In G14 the relation between true and observed mass has a gradient, leading to smaller masses
being systematically under-predicted by the model. Red galaxies in G14 tend to be affected by this more than blue galaxies. In L14 the differences in mass
have a larger scatter, the blue galaxies still show a gradient but the inferred mass is now generally an over-prediction of the true mass. The scatter between the
inferred and true mass is also larger for the L14 model.
Applying the colour relation to estimate mass, as compared to using the true mass, can cause both the masses and the typical colours of galaxies to change
in the sample. \citet{campbell15} demonstrate these differences can affect the clustering signal, with the effect being particularly marked for the L14 model. In
this paper we take the model masses at face value, except for a small adjustment explained in Sec.~\ref{sec:modelcuts}. We discuss the possible effects of this when 
interpreting results.

\subsubsection{Literature {\sc galform} comparisons}
The {\sc galform} model has been compared to a variety of observational measurements, including those made from this GAMA sample. Indeed several of 
the {\sc galform} models were created to address a disagreement with observations. For example, \citet{kim09} compared the \citet{bower06} and 
\citet{font08} models to 2dFGRS data and found that the model predictions for the dependence of
clustering on luminosity were not successfully reproduced. \citet{kim09} also found that the model had excess small scale clustering. The authors 
put this down to the \citet{bower06} model having too many satellite galaxies. 

In support of this, \citet{tamsyn14} compares the luminosity function as a function of environment of the \citet{bower06} mocks to those of 
GAMA galaxies, and finds an excess of faint, red galaxies in the model. \citet{robotham11} additionally found an excess of high-multiplicity
(10 or more galaxies) groups in the \citet{bower06} model compared to the GAMA data. 

\citet{campbell15} combine SDSS and VIPERS measurements with the ones presented here to compare the clustering of galaxies as a function of stellar mass to that 
predicted by a variety of {\sc galform} versions, across several different epochs. They find good agreement between the models and data, but in order to fit small scale clustering they have to adopt a 
new, hybrid model for satellite orbits. In this model satellites follow resolved sub haloes and an analytic merger timescale is only computed once 
the subhalo is lost. 

Their paper focuses on how different methods of estimating stellar mass affect clustering measurements; our paper takes the 
complementary approach of comparing a wider range of galaxy properties. In addition, our paper utilises the
lightcone modelling of \citet{merson13}, whilst \citet{campbell15} uses clustering of galaxies in a single, constant cosmic time, snapshot.

\subsection{Sample Selection}\label{sec:samp}

\begin{figure*}
 \noindent
\includegraphics[width=1.0\textwidth]{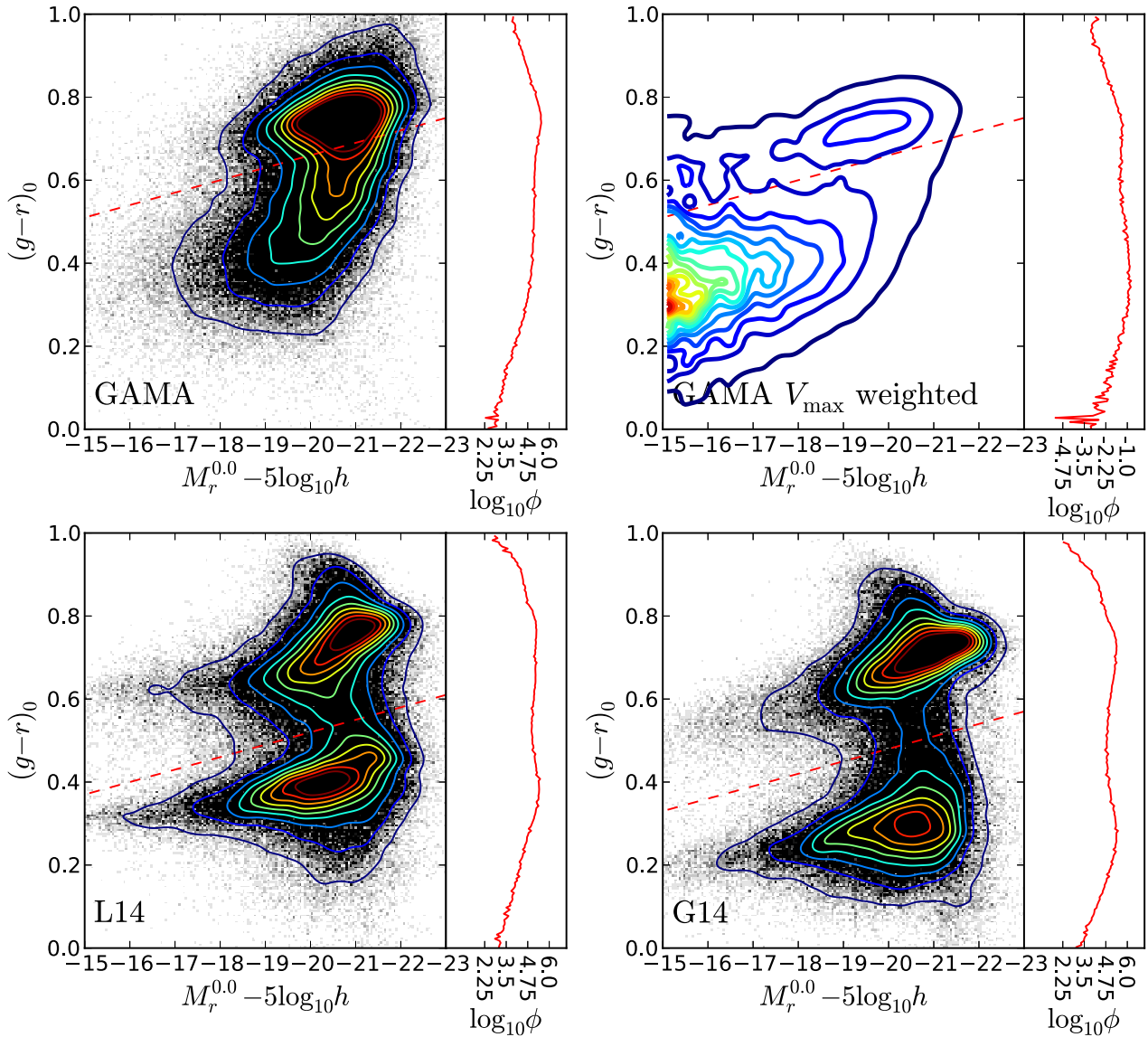}
 \caption{A restframe colour-magnitude diagram for the real GAMA galaxies and for the mock galaxies, as labelled. The top right panel gives the 
volume-weighted colour-magnitude diagram for the real GAMA galaxies. The red dashed lines shows our cuts to define red and blue samples of galaxies,
this cut is different for each model and the data (see Section~\ref{sec:modelcuts}). A 
histogram of colours has been added to the right of each panel. Contour levels are spaced linearly, the colour scale is the same in each of the
colour-magnitude diagrams, except the volume weighted one. The contours were computed from a binned version of this plot, smoothed with a 
Gaussian filter.} \label{fig:gamaCMD}
\end{figure*}
\subsubsection{Redshift slices}
We want to study the evolution of galaxy clustering with luminosity, mass, colour and redshift. One approach to this is to use volume limited samples,
which are characterised by a uniform detection probability across the sample volume. However, volume limited samples reduce the amount of data available for 
the analyses, so we also consider magnitude-limited samples. In order to do this, we ensure the survey's radial selection function is properly dealt with by 
the random catalogue (see Section \ref{sec:randGAMA}). Using magnitude-limited samples means interpreting the results requires consideration of the selection function, but offers the most powerful test of the model. We separate all of our mass, luminosity and colour samples into four redshift bins: low-z, $0.02<z<0.14$, intermediate-z, $0.14<z<0.24$, upper intermediate-z,  $0.24<z<0.35$ and high-z, $0.35<z<0.5$. The volumes of these slices are 
$1.2 \times 10^6$~(Mpc/$h)^3$, $5.0 \times 10^6$~(Mpc/$h)^3$, $1.3 \times 10^{7}$~(Mpc/$h)^3$ and $3.7 \times 10^{7}$~(Mpc/$h)^3$ respectively. 

In order to help differentiate between selection effects and galaxy formation effects, we also produce samples with upper redshift cuts that have 
been decreased in order to make the sample essentially volume limited (comprise of at least 98\% of galaxies with maximum observable redshifts greater 
than the sample limit).\footnote[3]{Applying the same cuts to the mock catalogues leads to a larger fraction of some mock samples being magnitude limited, at worse 
the volume-limited fraction is 89\%. Clustering from a mock catalogue for samples even more magnitude-limited than this (as low as 78\%) show no significant differences 
with the clustering from the same mock with a deeper apparent magnitude cut ($r<21$). We will therefore treat the mock measurements as volume limited. }

\subsubsection{Luminosity, mass and colour cuts}\label{sec:samplmc}
Luminosity samples are produced from Eq.~\ref{eq:mtoM} with the measured SDSS $r$-band DR7 Petrosian magnitude used to calculate the absolute magnitude 
in the $r$-band, $M_{r,h}$. \citet{loveday15} measures the luminosity function, and finds the characteristic ``knee'' of the function at 
$M_{r,h}^* =-20.6$; past this magnitude the number density of galaxies rapidly drops. This characteristic magnitude lies in our 
$-21.0<M_{r,h}<-20.0$ sample, and we measure one sample of galaxies brighter than this and three fainter.

Stellar mass is more complex to compute from observations, but is less complicated to predict from models. We use mass bins of $0.5$~dex in size, except
for the lowest mass sample, where we increase the bin limits to have more galaxies. \citet{baldry12} found for GAMA data at $z<0.06$ that the 
characteristic knee of the mass function is at $M^* = 10^{10.35} \rm{\mdot} /h^{2}$. This characteristic mass falls into our middle mass bin, we 
use two samples less massive and two samples more massive than this.

In addition to luminosity and mass samples, galaxies are divided into a red population and a blue population. The colour of a galaxy is often used as 
a rough proxy for the age of its stellar population, with galaxies undergoing star formation generally being expected to be bluer. It is therefore 
interesting to study how clustering differs as a function of colour. In Fig.~\ref{fig:gamaCMD} we show the colour-magnitude diagram for our flux limited
($r<19.8$) sample of galaxies. We use a sloping cut with a gradient of 0.03, following the typical literature values \citep[e.g.][]{bell03}, 
\begin{equation}
(g-r)_{0} = -0.030*(M_{r,h} - M_{r,h}^*) + 0.678
\end{equation}
The intercept of this cut is such that there is an equal fraction of red and blue galaxies. We also show in Fig.~\ref{fig:gamaCMD} a volume-weighted 
colour-magnitude diagram. We see that the volume we observe has an obvious blue cloud, made up of faint galaxies, which is less apparent in the observed colour-magnitude diagram
due to selection effects. The colour cut also separates the red sequence and blue cloud in the volume-weighted diagram.  

A summary of our cuts and the properties of our samples is given in Table~\ref{table:gamaSamps1}. 

\subsubsection{Cuts and adjustments to the models}\label{sec:modelcuts}
When comparing the actual clustering of galaxies to models we want to uncover differences in how dark matter haloes are populated. If we were 
to apply the same cuts on the models and the data we may get samples with very different number densities, since the luminosities or colours may be 
incorrectly predicted. Such disagreements are better explored by comparison of model predictions to other observational measurements like stellar mass 
functions, or colour distributions. In order to focus on the spatial distribution of comparable galaxies in the models and simulations, mock galaxy 
samples are often selected to have matching number densities to galaxies in the real comparison sample \citep[e.g][]{berlind03, zheng05, contreras13}. We
therefore adjust the magnitudes of the mock galaxies in order for the luminosity functions of the mocks and data to match, in each of our four redshift
slices.

First we estimate the galaxy luminosity function of the $r<21$ unaltered mock magnitudes. We then reassign each galaxy in the mock a magnitude designed
to map number densities in the unaltered mock luminosity function to the appropriate magnitude in the data luminosity function. We do this by computing
the magnitude difference between the cumulative mock luminosity function and the cumulative data luminosity function at the same cumulative number density. This forces
 the mock and data luminosity functions to agree. The faint end of the real luminosity functions suffer from incompleteness, as the different k-corrections 
of the galaxies result in a colour dependence on which galaxies reach the apparent magnitude limit. In order to avoid this we estimate the complete  
luminosity function of real data in the three higher redshift slices by switching to the low-redshift real luminosity function past the point where 
colour-dependent completeness effects become important. The amplitude of the low-redshift luminosity function is scaled to ensure a smooth join with the 
higher redshift one. The mock catalogue then has an $r<19.8$ apparent magnitude limit applied to the shifted magnitudes. 

We apply the same k-correction polynomials to the mocks, but adjust the limits of the colour bins. This is to ensure the fraction of galaxies in each 
colour bin is the same as in the data. Thanks to this adjustment, applying the apparent magnitude limit results in a mock catalogue with close to the same selection function as the 
real data. This is important as many of our samples are magnitude limited. Unfortunately, unless the mocks have the same colour-magnitude distribution as the data, it is impossible to perfectly model the selection function even with 
this technique of adjusting the mock k-corrections. This is because the magnitude distribution within a colour bin varies between real and mock data. The masses of the 
mock galaxies are rescaled to preserve the same mass-to-light ratio. Also note the mock catalogue is fit with the same $P$ and $Q$ as the data.

We will now describe the cuts for colour selected samples. In Fig.~\ref{fig:gamaCMD} we can see that both models reproduce the bimodality of galaxy 
colours, but both predict a well-defined blue cloud not seen in the data. Note that whilst the data is flux limited, meaning faint galaxies may be lost,
this selection effect is included in the model lightcone. One suggestion of how to improve the mock galaxy colours was given in
\citet{font08}; here the authors removed the unrealistic {\sc galform} assumption that all gas is removed from satellites as soon as they accrete onto
a halo. By implementing a more realistic model with more gradual removal of gas by ram pressure stripping they found satellites
could form stars for longer after being accreted, and ended up with colours between the red and blue sequence. However, the G14 and L14 models do not include any 
gradual ram pressure stripping.

Looking at the models in more detail, we see the L14 has a redder blue cloud than the G14 model. This is expected as this model uses the
\citet{maraston05} SPS model which has stronger near-IR emission for young stellar populations. We use the same gradient as the data for the cut, and select the 
intercept in order to reproduce the observed fraction of red and blue galaxies. For the L14 model this intercept it 0.548, for the G14 model it is 0.498. From 
Fig.~\ref{fig:gamaCMD} we can see that the cut applied to the models clearly separates the red sequence from the blue-sequence mock galaxies.

In Table~\ref{table:g14Samp} and Table~\ref{table:l14Samp} in Appendix~\ref{sec:tables}, the sample sizes and properties for one realisation of GAMA are given for 
the two models.

\subsubsection{Comparison samples}\label{sec:compSamp}
In addition to the samples we have described, we also produce a sample with which to compare our results to \citet{zehavi11}. In order to make the comparison 
samples as similar as possible, we use magnitudes corrected to $z_{\rm{ref}}=0.1$ and use the redshift cuts stipulated in \citet{zehavi11}. These magnitudes will be labelled 
with the superscript `0.1'. Unfortunately, the \citet{zehavi11} redshift cuts greatly restrict the volume of our survey. As such, only two of 
the \citet{zehavi11} magnitude bin samples,  $-22.0<M^{0.1}_{r,h}<-21.0$ and $-23.0<M^{0.1}_{r,h}<-22.0$ have a large enough volume in GAMA for worthwhile 
comparison, having volumes of $6.2 \times 10^{6}$~(Mpc$/h)^{-3}$ and $1.6 \times 10^{6}$~(Mpc$/h)^{-3}$ respectively. The redshift cuts for the fainter 
magnitude bins result in extremely small volumes in GAMA (less than $1.1 \times 10^{5}$~(Mpc$/h)^{-3}$).

We also produce a series of volume-limited samples with which to compare to SDSS, using deeper redshift cuts appropriate to GAMA's fainter selection limit. These
samples are included in Table~\ref{table:gamaSamps1}. 

\begin{table*}
\begin{center}
  \caption{Different galaxy samples, sample sizes and median properties. The subscript `med' indicates the values are medians. 
   Samples with redshift limits marked with asterisks are magnitude limited, so should be treated with careful consideration
   of the GAMA selection function. In samples without an asterisk at least 98\% of the members are volume limited. Values in brackets 
   are rms scatter.}
  \medskip
  \begin{tabular}{c c c c c c c c c}
    \hline\hline
    Sample  & $z_{\mathrm{min}}$  & $z_{\mathrm{max}}$ & $N_{\rm{gals}}$ &   $z_{\mathrm{med}}$  & $M_{\mathrm{med}}$ & $\log_{10}(M_{*}/\mathrm{\mdot}h^{-2})_{\mathrm{med}}$ & $(g-r)_{\mathrm{0, med}}$ \\
    \hline
    \hline
    $-18.00<M_{r,h}<-17.00$ & 0.02 & 0.07 & 2089 &  0.05 & -17.46 (0.28) & 8.62 (0.29) & 0.42 (0.20) \\
    $$ & 0.02 & 0.14* & 5666 &  0.08 & $-17.62$ (0.27) & 8.67 (0.33) & 0.41 (0.19)\\
    \hline
    $-19.00<M_{r,h}<-18.00$ & 0.02 & 0.11 & 6950 &  0.09 & -18.47 (0.29) & 9.13 (0.31) & 0.47 (0.21) \\
    $$ & 0.02 & 0.14* & 13149 &  0.11 & $-18.54$ (0.27) & 9.15 (0.31) & 0.47 (0.19)\\
    $$ & 0.14 & 0.24* & 3307 &  0.15 & $-18.85$ (0.13) & 9.26 (0.28) & 0.46 (0.16)\\
    \hline
    $-20.00<M_{r,h}<-19.00$ & 0.02 & 0.14 & 11741 &  0.11 & $-19.47$ (0.29) & 9.74 (0.29) & 0.62 (0.17)\\
    $$ & 0.14 & 0.17 & 7801 &  0.16 & -19.46 (0.29) & 9.74 (0.31) & 0.60 (0.22) \\
    $$ & 0.14 & 0.24* & 27755 &  0.18 & $-19.59$ (0.27) & 9.79 (0.32) & 0.59 (0.20)\\
    $$ & 0.24 & 0.35* & 2834 &  0.26 & $-19.90$ (0.10) & 9.81 (0.38) & 0.45 (0.18)\\
    \hline
    $-21.00<M_{r,h}<-20.00$ & 0.02 & 0.14 & 6945 &  0.11 & $-20.40$ (0.28) & 10.24 (0.25) & 0.71 (0.15)\\
    $$ & 0.14 & 0.24 & 22327 &  0.20 & $-20.39$ (0.28) & 10.27 (0.28) & 0.70 (0.15)\\
    $$ & 0.24 & 0.35* & 37541 &  0.28 & $-20.55$ (0.27) & 10.39 (0.28) & 0.68 (0.17)\\
    $$ & 0.35 & 0.50* & 3593 &  0.37 & $-20.86$ (0.14) & 10.42 (0.28) & 0.55 (0.16)\\
    \hline
    $-22.00<M_{r,h}<-21.00$ & 0.02 & 0.14 & 1843 &  0.12 & $-21.27$ (0.25) & 10.67 (0.19) & 0.74 (0.11)\\
    $$ & 0.14 & 0.24 & 5865 &  0.20 & $-21.28$ (0.25) & 10.70 (0.21) & 0.74 (0.13)\\
    $$ & 0.24 & 0.35 & 15353 &  0.30 & $-21.29$ (0.25) & 10.78 (0.23) & 0.74 (0.17)\\
    $$ & 0.35 & 0.37 & 3168 &  0.36 & -21.28 (0.25) & 10.82 (0.23) & 0.74 (0.14)\\
    $$ & 0.35 & 0.50* & 16114 &  0.40 & $-21.42$ (0.27) & 10.86 (0.28) & 0.73 (0.22)\\
    \hline
    \hline
    $-23.00<M^{0.1}_{r,h}<-22.00$ & 0.01 & 0.50 & 2273 &  0.37 & $-22.25$ (0.19) & 11.26 (0.28) & 0.79 (0.18)\\
    $-22.00<M^{0.1}_{r,h}<-21.00$ & 0.01 & 0.38 & 23746 &  0.29 & $-21.36$ (0.25) & 10.79 (0.23) & 0.74 (0.15)\\
    $-21.00<M^{0.1}_{r,h}<-20.00$ & 0.01 & 0.26 & 33965 &  0.20 & $-20.47$ (0.28) & 10.31 (0.28) & 0.71 (0.16)\\
    $-20.00<M^{0.1}_{r,h}<-19.00$ & 0.01 & 0.18 & 22487 &  0.14 & $-19.53$ (0.29) & 9.78 (0.31) & 0.62 (0.19)\\
    $-19.00<M^{0.1}_{r,h}<-18.00$ & 0.01 & 0.12 & 8129 &  0.09 & $-18.52$ (0.29) & 9.16 (0.31) & 0.48 (0.21)\\
    \hline
    \hline
    $8.50<\log_{10}M_{*}/\mathrm{\mdot}h^{-2}<9.5$ & 0.02 & 0.05 & 1049 &  0.04 & -17.95 (0.72) & 8.88 (0.28) & 0.46 (0.15) \\
    $$ & 0.02 & 0.14* & 18533 &  0.10 & $-18.42$ (0.59) & 9.07 (0.27) & 0.44 (0.13)\\
    $$ & 0.14 & 0.24* & 8472 &  0.17 & $-19.19$ (0.33) & 9.32 (0.16) & 0.39 (0.10)\\
    \hline
    $9.50<\log_{10}M_{*}/\mathrm{\mdot}h^{-2}<10.00$ & 0.02 & 0.14 & 10053 &  0.11 & $-19.43$ (0.47) & 9.74 (0.14) & 0.66 (0.12)\\
    $$ & 0.14 & 0.24* & 19173 &  0.18 & $-19.64$ (0.37) & 9.79 (0.14) & 0.56 (0.12)\\
    $$ & 0.24 & 0.35* & 5331 &  0.27 & $-20.11$ (0.25) & 9.84 (0.13) & 0.41 (0.11)\\
    \hline
    $10.00<\log_{10}M_{*}/\mathrm{\mdot}h^{-2}<10.50$ & 0.02 & 0.14 & 7385 &  0.11 & $-20.28$ (0.44) & 10.21 (0.14) & 0.73 (0.12)\\
    $$ & 0.14 & 0.18 & 7749 &  0.16 & -20.25 (0.43) & 10.21 (0.14) & 0.73 (0.09) \\
    $$ & 0.14 & 0.24* & 23216 &  0.20 & $-20.24$ (0.40) & 10.22 (0.14) & 0.72 (0.10)\\
    $$ & 0.24 & 0.35* & 23584 &  0.28 & $-20.48$ (0.29) & 10.32 (0.14) & 0.63 (0.12)\\
    $$ & 0.35 & 0.50* & 3132 &  0.38 &  $-20.92$ (0.26) & 10.36 (0.13) & 0.48 (0.10)\\
    \hline
    $10.50<\log_{10}M_{*}/\mathrm{\mdot}h^{-2}<11.00$ & 0.02 & 0.14 & 2189 &  0.12 & $-21.15$ (0.38) & 10.65 (0.12) & 0.76 (0.10)\\
    $$ & 0.14 & 0.24 & 8018 &  0.20 & $-21.10$ (0.38) & 10.65 (0.13) & 0.76 (0.09)\\
    $$ & 0.24 & 0.29 & 9968 &  0.27 & -21.00 (0.38) & 10.66 (0.13) & 0.76 (0.09) \\
    $$ & 0.24 & 0.35* & 24416 &  0.30 & $-21.02$ (0.35) & 10.68 (0.13) & 0.75 (0.09)\\
    $$ & 0.35 & 0.50* & 11800 &  0.38 & $-21.31$ (0.26) & 10.78 (0.14) & 0.69 (0.12)\\
    \hline
    $11.00<\log_{10}M_{*}/\mathrm{\mdot}h^{-2}<11.50$ & 0.24 & 0.35 & 2572 &  0.31 & $-21.86$ (0.38) & 11.10 (0.10) & 0.80 (0.15)\\
    $$ & 0.35 & 0.37 & 698 &  0.36 & -21.74 (0.35) & 11.11 (0.11) & 0.82 (0.12) \\
    $$ & 0.35 & 0.50* & 5962 &  0.41 & $-21.81$ (0.29) & 11.14 (0.12) & 0.82 (0.11)\\
    \hline
    \hline
    $\rm{Red}$ & 0.02 & 0.14* & 15842 &  0.11 & $-19.61$ (1.12) & 9.98 (0.54) & 0.73 (0.17)\\
    $(g-r)_{0} + 0.03(M_{r,h} - M_{r,h}^*) > 0.678$ & 0.14 & 0.24* & 29686 &  0.19 & $-20.16$ (0.70) & 10.26 (0.35) & 0.75 (0.14)\\
    $$ & 0.24 & 0.35* & 30465 &  0.29 & $-20.81$ (0.50) & 10.62 (0.29) & 0.76 (0.15)\\
    $$ & 0.35 & 0.50* & 11461 &  0.39 & $-21.51$ (0.39) & 11.00 (0.29) & 0.80 (0.23)\\
    \hline
    $\rm{Blue}$ & 0.02 & 0.14* & 26437 &  0.10 & $-18.61$ (1.21) & 9.11 (0.62) & 0.44 (0.12)\\
    $(g-r)_{0} + 0.03(M_{r,h} - M_{r,h}^*) < 0.678 $ & 0.14 & 0.24* & 29906 &  0.19 & $-19.77$ (0.67) & 9.74 (0.42) & 0.51 (0.11)\\
    $$ & 0.24 & 0.35* & 26227 &  0.29 & $-20.56$ (0.49) & 10.26 (0.35) & 0.56 (0.12)\\
    $$ & 0.35 & 0.50* & 10122 &  0.39 & $-21.22$ (0.39) & 10.61 (0.30) & 0.58 (0.12)\\

\end{tabular}\label{table:gamaSamps1}
\end{center}
\end{table*}

\section{Methodology}\label{sec:methodGAMA}
In this section we will introduce our method to produce random catalogues, before explaining how we compute the 2PCF, $w_{p}(r_{p})$.

\subsection{Random catalogues}\label{sec:randGAMA}
\subsubsection{The Cole (2011) method}
To measure clustering one needs a random set of points with the same radial and angular selection function as the data. We 
generate catalogues of random positions using the method set out in \citet{cole11}, which generates random catalogues from the real data in a way that removes the effects
of the large-scale structure. We use the \citet{cole11} method without the complications of the extended method designed to fit a parameterised
evolution model to the galaxies. For each galaxy in the catalogue the maximum volume of space over which it could be 
observed, $V_{\rm{max}}$, is calculated by finding $z_{\rm{min}}$ and $z_{\rm{max}}$, the redshift where a galaxy meets the bright and faint magnitude 
limits of GAMA II. In addition to $V_{\rm{max}}$, a density weighted maximum volume, $V_{\rm{max,dc}}$, is calculated as
\begin{equation}\label{eq:vmaxdc}
V_{\rm{max,dc}} = \int_{z_{\rm{min}}}^{z_{\rm{max}}} \Delta(z) \frac{\mathrm{d}V}{\mathrm{d}z} \mathrm{d}z
\end{equation}  
where $\Delta(z)$ is the over-density as a function of redshift and $dV/dz$ is the comoving volume element per redshift element. Given these volumes, 
every real galaxy in the catalogue is cloned $n$ times, where $n$ is given by
\begin{equation}
n = n_{\rm{clones}} \frac{V_{\rm{max}}}{V_{\rm{max,dc}}}
\end{equation}
with $n_{\rm{clones}}$ being the total number of randoms divided by the number of galaxies in the sample. Our default random catalogues have $n_{\rm{clones}}=400$. When cloning the 
galaxy, all of the intrinsic galaxy properties are also cloned, so that the random points have a stellar mass, an absolute magnitude and a colour. The cloned galaxies are randomly distributed within the real galaxy's $V_{\rm{max}}$, with the GAMA angular mask used to ensure the angular selection 
function of the cloned galaxies matches that of the real galaxies. 

This method requires the estimation of $\Delta(z)$, which is done using an iterative method. Initially it is assumed $\Delta(z)=1$ everywhere such
that each galaxy is cloned the same number of times. From this random catalogue, $\Delta(z)$ is estimated from the redshift distribution of the 
randoms, $n_{\rm{r}}(z)$, and the data, $n_{\rm{g}}(z)$, using
\begin{equation}
\label{eq:over}
\Delta(z) = n_{\rm{clones}} \frac{n_{\rm{g}}(z)}{n_{\rm{r}}(z)}\, .
\end{equation}
A new random catalogue is then produced with this new estimate of $\Delta(z)$, and the whole process repeated until $\Delta(z)$ converges. We iterate 15 
times, and find this is more than enough iterations for the process to converge.  

An added bonus of this method is that the $V_{\rm{max,dc}}$ can be used to easily estimate the luminosity function \citep{cole11}. We use this technique when estimating luminosity
functions for the purpose of adjusting mock galaxy magnitudes to match the abundance of real galaxies. A study of the luminosity functions of GAMA galaxies from this estimator
is given in \citet{loveday15}. 
\subsubsection{Windowed Clones}
As previously mentioned, the creation of this random catalogue requires that one adopt a certain value of $Q$ (Eq.~\ref{eq:mtoM}). Additionally, 
our choice of $Q$ is motivated by our decision not to consider evolution in the overall number density of galaxies. It may be the case that our
chosen $Q$ is not true to the data, or some density evolution is present in the sample. To mitigate possible effects from this we add a new 
technique to the \citet{cole11} method.

The key idea of this method is to restrict the redshift of cloned galaxies to some window function around the redshift of the original galaxy. Evolution is 
therefore included naturally in the random catalogue, as cloned galaxies are kept close to their original redshift. Hence the error 
introduced by the inadequacy of the adopted $P$ and $Q$ model is limited. We choose to define our window as a function of volume, $W(V)$; this is 
because volume is most closely related to the expected fluctuations in the galaxy $\Delta(z)$. With a window defined in terms of volume, the $n(z)$ is 
smeared out more in the sample variance prone, low-redshift part of the $n(z)$ and less at higher redshift. We
adopt a Gaussian window function, with $\sigma=3.5 \times 10^{6}$(Mpc/$h$)$^3$, truncated at $2\sigma$. We found this to be a good compromise between 
limiting the effects of evolution and smoothing out large-scale structure. We also include this window in the computation of $V_{\rm max}$ and 
$V_{\rm max,dc}$, by weighting volume slices by the window when numerically computing the integral. When the window function reaches a boundary, either 
the $V_{\rm max}$ of an object or the limit of the survey, it is reflected from that boundary in order to ensure the randoms are scattered symmetrically 
across the volume. Simply truncating the window at a redshift limit would result in randoms being artificially more likely to be moved away from the 
limit. This method will be further explored, along with the choices of $\sigma$, in a future paper. We will, however, demonstrate one success of the method 
here empirically.
\begin{figure}
  \noindent
  \includegraphics[width=0.50\textwidth]{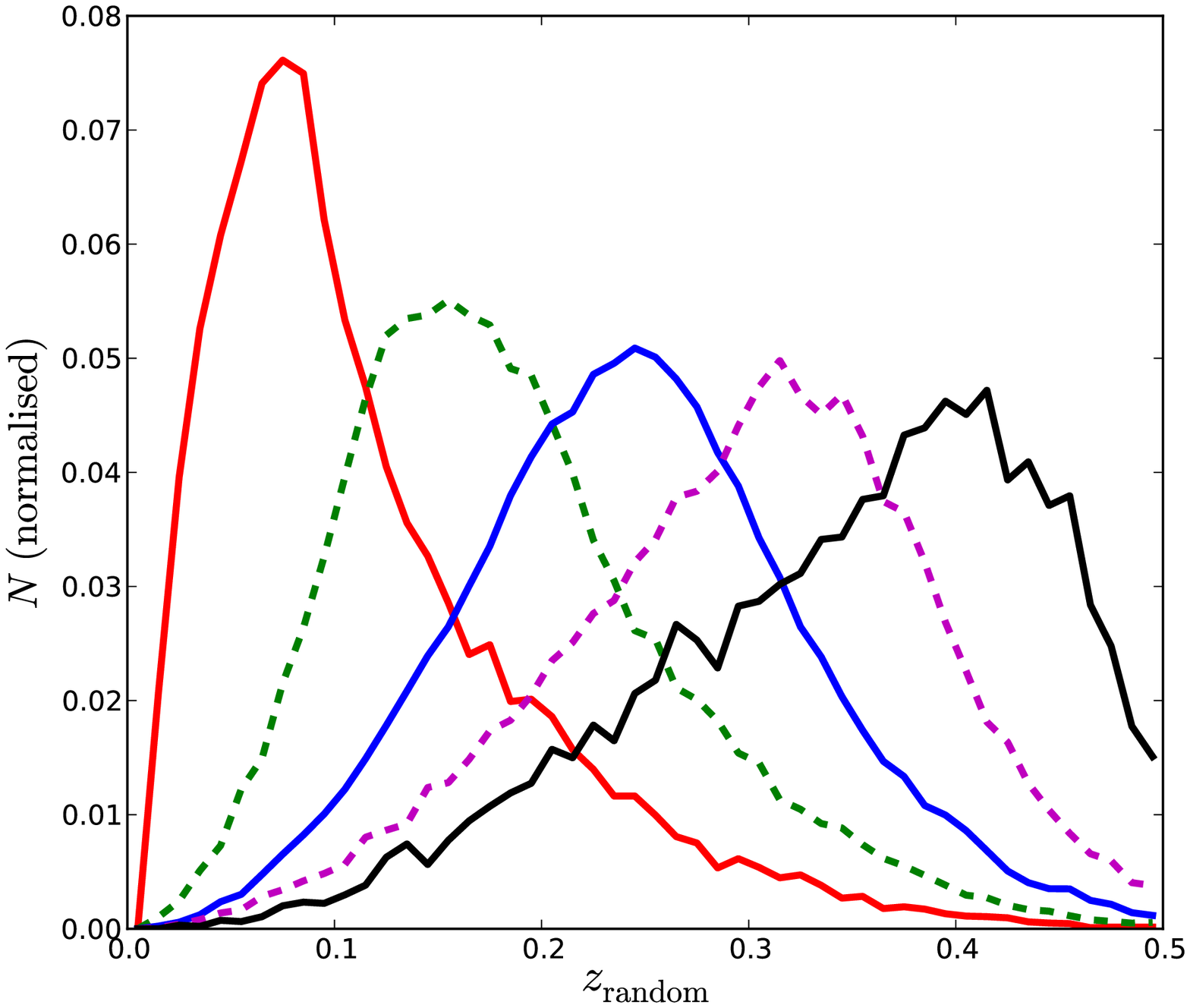}
  \includegraphics[width=0.50\textwidth]{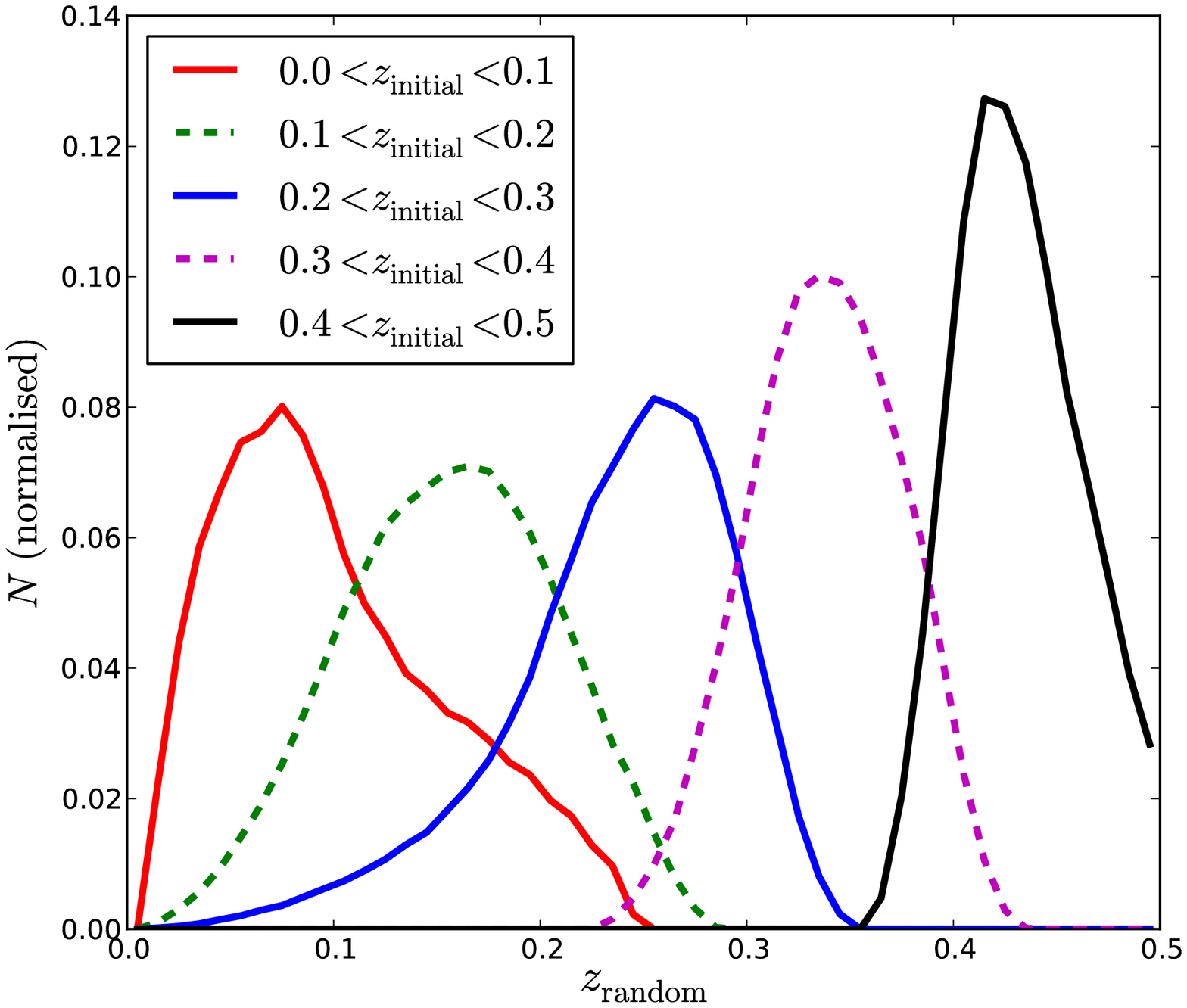}
  \caption{The distribution of the final redshifts of the random points, in bins of their initial redshift, for a random catalogue for the real GAMA data, 
    generated without (top) and with (bottom) the window function update. The window function update acts to limit how far a 
    cloned galaxy can move from its initial position. This in turn limits the effects of any unmodelled galaxy evolution on
    the random catalogue.}\label{fig:zintZfinal}
\end{figure}
In Fig.~\ref{fig:zintZfinal} we show the redshift distribution of GAMA cloned galaxies, for different bins of the original galaxy redshift. In the top
panel of Fig.~\ref{fig:zintZfinal} we can see that cloned galaxies are spread widely across the full redshift range of the survey. In the bottom
panel of Fig.~\ref{fig:zintZfinal} we can see that the addition of a window function has limited the redshift range over which a cloned galaxy can move. As such 
it has limited the effects of deviations of the evolution away from the description given by the adopted values of $P$ and $Q$. 

This windowing method requires further testing and tuning, to see how successful it is in removing the effects on the random catalogue from unmodelled
galaxy evolution. However, for this work we show in Appendix~\ref{sec:cTest} that the windowing affects our results much less than sample variance.

\subsubsection{The resultant catalogue}
\begin{figure*}
\begin{center}
 \includegraphics[width=6.0in]{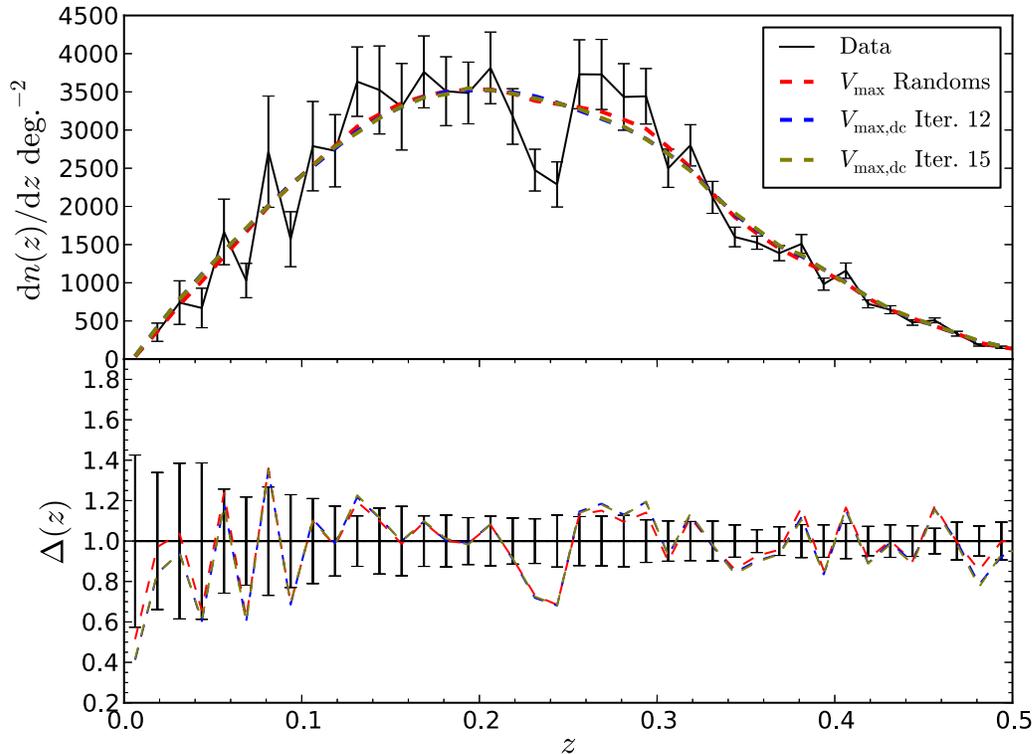}
 \caption{\emph{(Top)} The redshift distribution of GAMA II galaxies (solid) and our randoms (dashed) for multiple iterations of the Cole (2011) random catalogue 
   generating approach (see legend), combined with our window function modifications, as explained in Section 4.3.2. The redshift distribution of the randoms is a good match to the data. \emph{(Bottom)} Our estimates of 
   the galaxy over-density as a function of redshift, from the ratio of the galaxy and random redshift distributions. The error bars are from the rms scatter of the 26 mock realisations,
   as such they account for the large-scale structure variations. Note these error bars may be underestimates, as the mock realisations over-sample the simulation volume at high redshifts.} \label{fig:nzIter}
\end{center}
\end{figure*}

We show in Fig.~\ref{fig:nzIter}~(top) the redshift distribution of the data and the randoms for different iterations of this process. We see that 
using the density-corrected maximum volume (olive and blue dashed line) only introduces subtle differences into the $n_{\rm{r}}(z)$ of the randoms, as 
compared to simply using $V_{\rm{max}}$ (red dashed line). The twelfth iteration (blue) and the fifteenth iteration (olive) agree; this indicates the
process has converged. The $n_{\rm{r}}(z)$ of the randoms is a good fit to the $n_{\rm{g}}(z)$ of the total sample; later 
we will check the random $n_{\rm{r}}(z)$ is appropriate for galaxies split into magnitude, colour and mass samples. Note that it does appear that the
random catalogue slightly follows the under-density at $z \sim 0.22$; we found this was not the case when not using the window function and as 
such claim this is an unfortunate side effect of the modification to the method \citep{farrowThesis}. We have, however, tested our clustering results
using randoms from the old and new methods, and only find small differences that are at large scales. These differences are a fraction of the error bars, 
and as such our results will not be significantly affected. 

In Fig.~\ref{fig:nzIter}~(Bottom) we plot the over-density estimate, Eq.~\ref{eq:over}, for successive iterations of our random catalogue generating method. As expected, the iterations act to slightly increase 
the over-density estimates, as the \citet{cole11} method acts to remove their effect from the random catalogue. 
This method of generating randoms ameliorates the effects of large-scale structure on the generation of a random catalogue, as over-represented galaxies 
from over-dense regions are cloned fewer times whilst under-represented galaxies are cloned more times. \citet{cole11} demonstrates 
how this scheme can produce a random catalogue unbiased by large-scale structure. A strength of this approach to generating randoms is that the random catalogue 
comes with all of the properties of the galaxy catalogue. One can then apply the same selection to the random catalogues and the galaxy catalogues so that the 
random catalogue has the correct angular and radial distribution. We intend to make these random catalogues available to the GAMA community via a DMU, which
we also intend to distribute to the larger astronomical community once the GAMA data become public. 

\begin{figure}
  \noindent
  \includegraphics[width=0.50\textwidth]{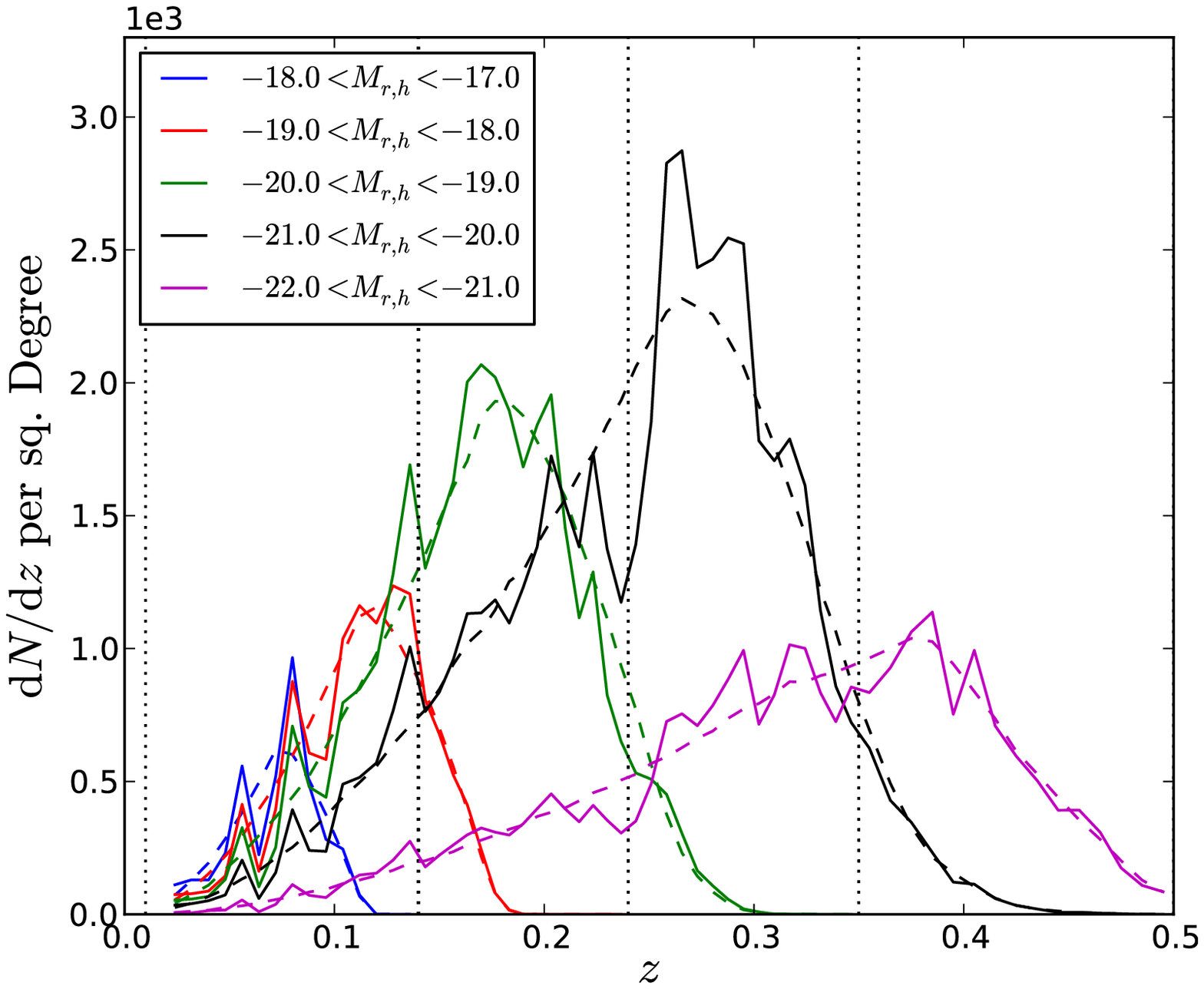}
  \includegraphics[width=0.50\textwidth]{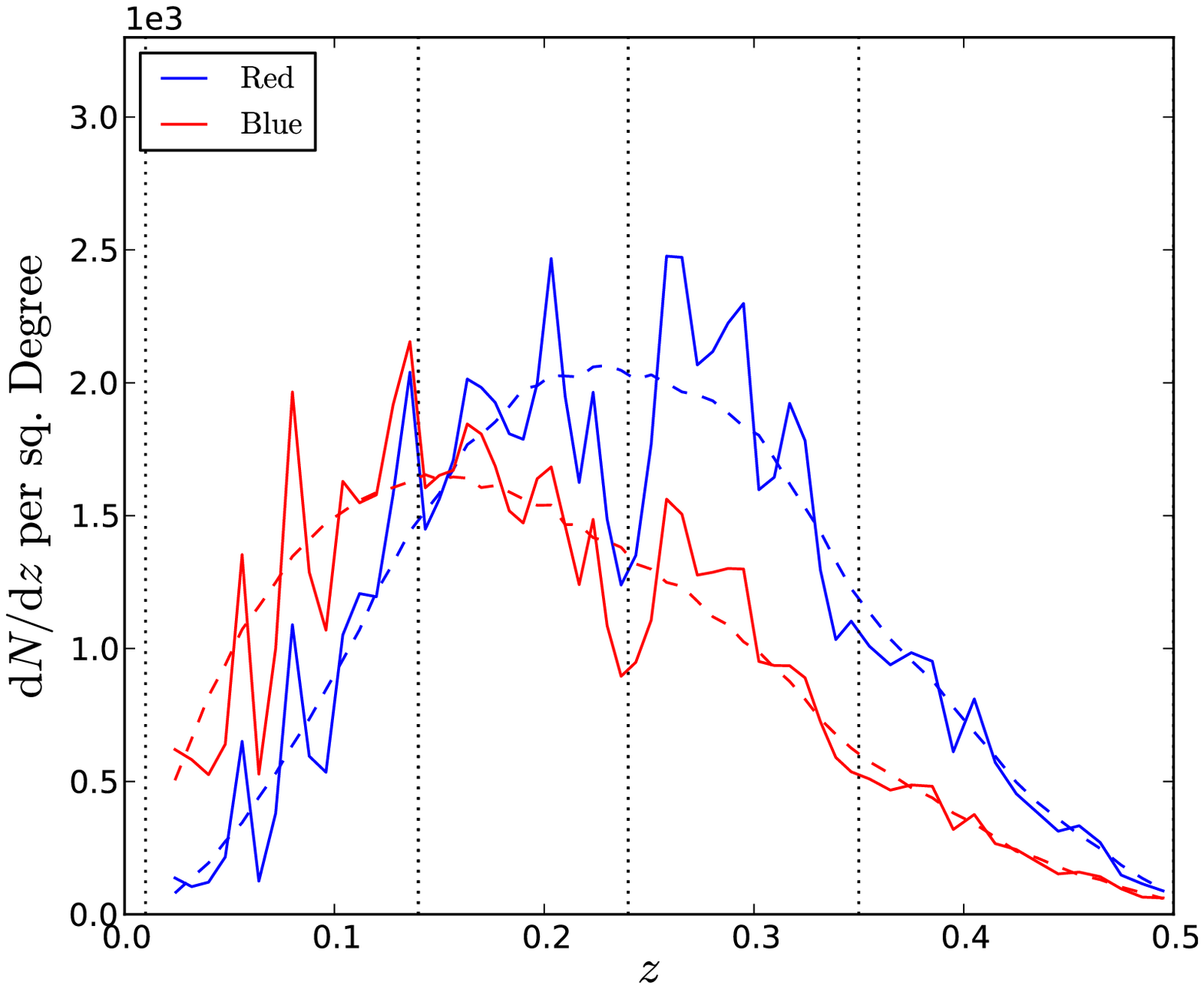}\\
  \includegraphics[width=0.50\textwidth]{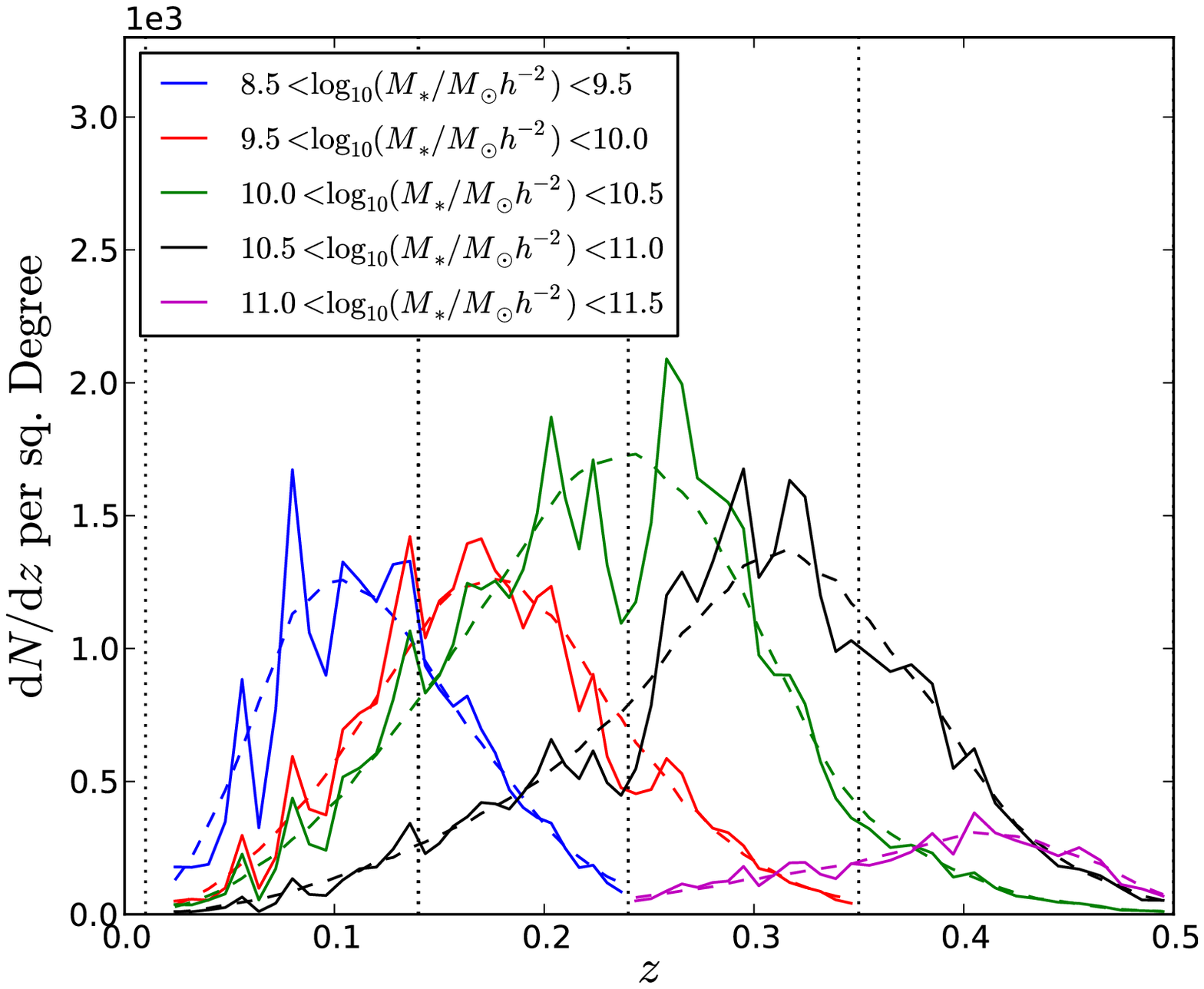}
  \caption{The redshift distribution of the data (solid lines) compared to the redshift distribution of the randoms (dashed lines), for different samples as indicated in the legend. Vertical dotted lines mark the positions of our redshift cuts. The randoms provide an excellent description of the data.}\label{fig:nzSamplesComp}
\end{figure}

In Fig.~\ref{fig:nzSamplesComp} we show the redshift distribution of the randoms and GAMA data, split into samples. We see for the luminosity, mass and colour samples we study the random redshift distribution is an excellent fit to the data.

\subsubsection{Randoms for the Mocks}
We apply the same techniques to generate randoms for the mock catalogues. This technique is applied first to the original mock catalogue with unadjusted magnitudes and 
$r<21.0$, in order to estimate the mock luminosity function. The technique is applied again to the mock catalogue with adjusted magnitudes and $r<19.8$, to yield the final random catalogue. We find that the resultant $n(z)$ of the randoms 
are a good fit to the mock catalogues.
\subsection{Projected clustering}\label{sec:proClus}
We measure the 2PCF of galaxies using pairs of galaxies and randoms with the \citet{hamilton} estimator. Following the 
standard approach adopted in the literature \citep[e.g.][]{coil08} we measure pair separations parallel, $\pi$, and transverse, $r_p$, to the line of 
sight for each pair. These are computed by first converting the angular position and redshift of each object to a vector, ${\bf r}$. We 
then define a line-of-sight direction to a pair as ${\bf l}=({\bf r_{1} + r_{2}})/2$, where ${\bf r_1}$ and ${\bf r_2}$ are the positions of the two pair 
members. The parallel to the line-of-sight distance, $\pi$, is the projection of the separation, ${\bf s} = {\bf r_2} - {\bf r_1}$, onto the line of sight
\begin{equation}
\pi = \frac{{\bf s}\cdot{\bf l}}{|{\bf l}|}    .
\end{equation}
The separation transverse to the line of sight is then
\begin{equation}
r_p = \sqrt{ |s|^2 - \pi^2} .
\end{equation}
Pairs are binned onto a grid of $\pi$ and $r_p$. In order to maximise signal-to-noise ratio we use a mixed linear and logarithmic binning scheme. Bins are linear with 0.0625~$h^{-1}$Mpc width up to 0.3~$h^{-1}$Mpc, and are 0.12~dex in size for bins larger than that. Whilst we generate 
400 times as many random points as the data, when computing clustering we use between $n_{\rm{clones}}=32$ and $n_{\rm{clones}}=200$ times more randoms than 
data, the number varies in order to maintain good statistics for the samples where the galaxies only sparsely populate the volume.

It is well known that the measured correlation function is distorted by the peculiar velocities of galaxies. On larger scales the infall 
of galaxies squash the observed correlation function in the line-of-sight direction \citep{kaiser87}. 
On small scales the virial motions of galaxies within clusters can elongate the correlation function along the line of sight \citep{jackson72}.  These distortions have been studied in the GAMA data by Loveday et al (in prep.)
and \citet{blake13}; in this work we focus instead on the projected correlation function, $w_p(r_p)$. The projected correlation function is a standard approach for dealing with redshift space distortions, which involves 
integrating $\xi(r_p, \pi)$ along the $\pi$ direction to minimise their effects, thus
\begin{equation}
\label{eq:intwprp}
w_p(r_p) = 2 \int_0^{\pi_{\rm{max}}} \xi(r_p, \pi) \rm{d}\pi .
\end{equation}
In practice this integral is carried out numerically using our $\xi(r_p, \pi)$ grid. The choice of $\pi_{\rm{max}}$ warrants careful consideration. Ideally,
one would use the largest possible value of $\pi_{\rm{max}}$ to include most of the 2D clustering signal, and because theoretically the
effects of redshift space distortions are only removed if you integrate out to infinity. Unfortunately, in real surveys noise affects the measurements,
and measurements for large values of $\pi_{\rm{max}}$ can be particularly noisy. We adopt a value of $\pi_{\rm{max}}=41\,{\rm Mpc}\,h^{-1}$ for our measurements, 
which is a reasonable $\pi_{\rm{max}}$ value to consider for the scales analysed  \citep[see e.g. Fig. 1 of][]{norberg09}. For the \citet{zehavi11} comparison sample
we use $\pi_{\rm{max}}=60\,{\rm Mpc}\,h^{-1}$, which makes our measurements noisier but matches the \citet{zehavi11} $\pi_{\rm{max}}$.

The results of Eq.~\ref{eq:intwprp}, with $\pi_{\rm{max}}=\infty$, can be calculated analytically for a spherically symmetric power law correlation function, 
$\xi(r_p)=(r_p/r_0)^{-\gamma}$, where $r_0$ and $\gamma$ are constants. The result is 
\begin{equation}
\label{eq:wprp}
w_p(r_{\mathrm{p}}) = r_p \left(\frac{r_0}{r_{\mathrm{p}}}\right)^{\gamma}\frac{\Gamma(1/2)\Gamma((\gamma -1)/2)}{\Gamma(\gamma/2)}
\end{equation} 
where $\Gamma$ is the Gamma Function. We will fit Eq.~\ref{eq:wprp} to some of our samples in Section~\ref{sec:concGAMA} in order to measure correlation 
lengths, $r_0$. As a power law is not a good fit over the whole 2PCF we restrict the fit to the scales  $0.2\,\mathrm{Mpc}\,h^{-1}<r_{p}<9.0\,\mathrm{Mpc}\,h^{-1}$. To fit 
we adopt a least-squares minimisation method using the diagonal terms of the covariance matrix. To check how the variation of $\gamma$ could influence the
derived value of $r_{0}$, we also fit power laws with a fixed $\gamma=-1.8$. All of the measured values are given in the Appendix in Table~$\ref{table:fits}$. We see only
a small difference between the best fitting $r_{0}$ values where $\gamma$ is free to vary and where $\gamma=-1.8$. The difference is not large enough to
affect our conclusions.

When plotting clustering, we often include a reference power law line or divide through by this reference power law, $w_{\rm{ref}}$, to allow easier 
comparison between plots. We use the \citet{zehavi11} power law fit to their $-21.0<M_{r,h}^{0.1}<-20.0$ sample for this purpose, which has 
$r_{0} = 5.33h^{-1}$~Mpc and $\gamma=-1.81$. 

The clustering results in this paper have been cross-checked numerous times to great accuracy against independent clustering analyses.

\subsection{Error estimates}\label{sec:jackknife}
To compute error bars on our clustering measurements for GAMA data, 9 jack-knife samples \citep[e.g.][]{zehavi02}, 3 per region, are formed by 
rejecting roughly equal-area regions of data. From this method, the covariance matrix $C_{ij}$, is calculated, the square-root of the diagonal terms 
of which give the error bars. Work such as \citet{guo14} show that with large surveys like SDSS, a large number of jack-knife resamplings
($\sim 100$) can give reliable estimates of the covariance matrix. Our smaller area, necessitating a smaller number of jack-knife regions, 
means we should further test our covariance matrix estimates.

\begin{figure*}
 \includegraphics[width=1.0\textwidth]{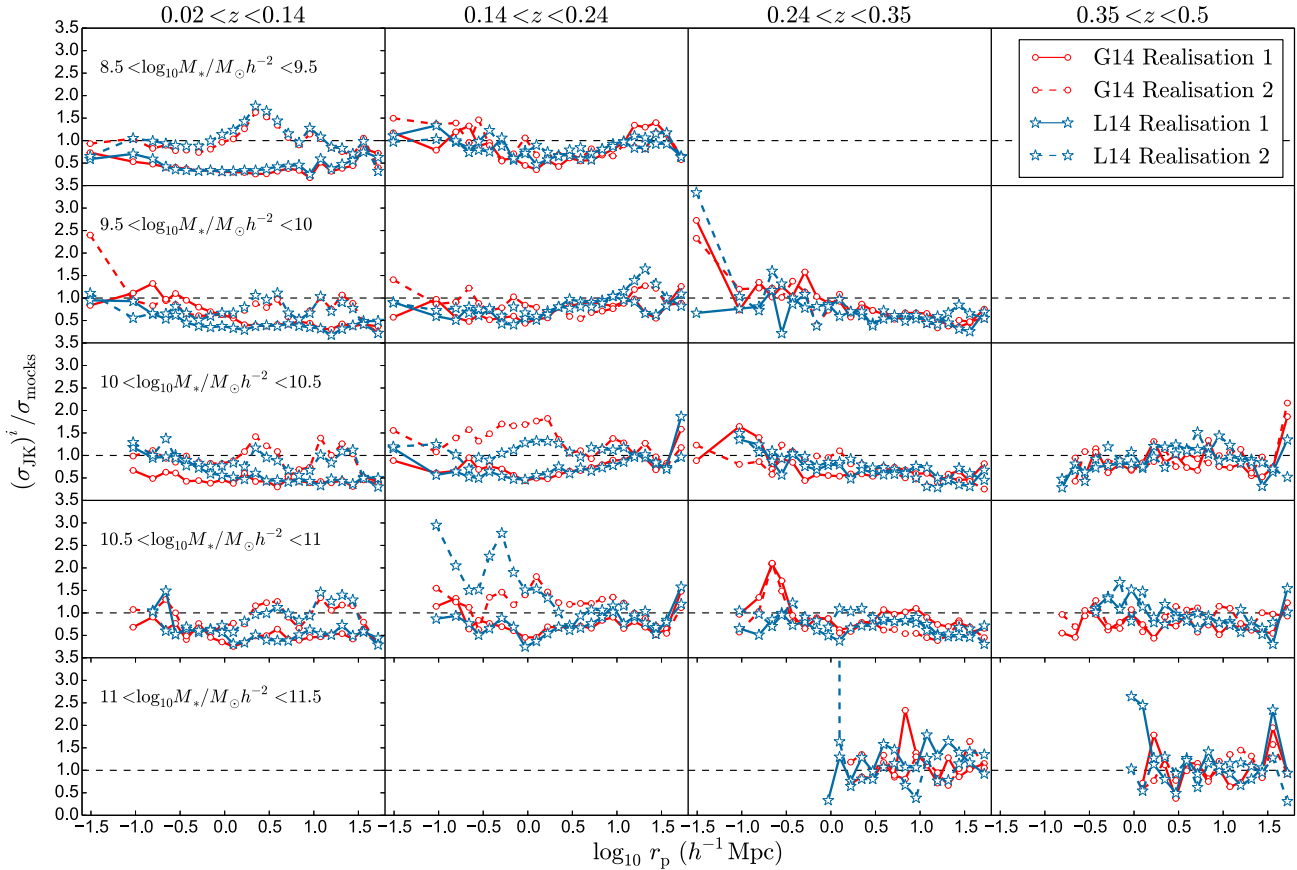}
 \caption{The ratio of error computed from the scatter between 26 mock realisations, and jack-knife error estimates from two different realisations. Red lines 
  are from the L14 model whilst blue lines are from the G14 model. We can see the error estimates from jack-knife resampling can be very different to those
  from the mock realisations.}\label{fig:mockComp}
\end{figure*}

We also test how unbiased using jack-knife errors is for our sample. For this we utilise the 26 realisations of each model, created by
considering different lightcones through the simulation. These estimates of the error should be more realistic, as they are built from a series of lightcones which 
together sample a larger volume than the GAMA data, even though they are drawn from a single simulation. This latter fact implies that the mock errors shown are 
likely underestimates of the true sample variance, as estimated from an series of lightcones constructed from independent simulations.
We computed the jack-knife errors, for a single mock, and the root mean square (rms) scatter between the different realisations for all of the samples selected on mass. 
In Fig.~\ref{fig:mockComp} we show the ratio of the two error estimates as a function of projected separation. Different colours are from the different models, as 
indicated. The dashed and solid lines represent two different realisations of the model, encompassing different volumes of the simulation. These two volumes are the same
for both models, in order to disentangle the effects of sample variance and galaxy formation.

In the highest redshift slices the error estimates converge, however note that this convergence could be due to the two realisations sampling some of the same volume of 
the N-body simulation (see Section~\ref{sec:lightcones}). In the two lower redshift slices the realisations should be nearly independent. We can see that for some redshift and mass 
ranges the jack-knife errors give the same uncertainty as the model, whilst for others the errors disagree. In particular, the low-redshift jack-knife errors give a lower estimate in 
 the first volume. This suggests for our samples jack-knife errors could be underestimates at low redshift. As expected, we only see a small difference between the ratios for the two 
 models in the same volume, as such it appears that the ratio is most sensitive to sample variance. 
Despite the limitations of the jack-knife error estimates with our samples, we use them with the data as it has the advantage of not being model dependent, nor sensitive to how accurately 
the models reproduce the observed GAMA clustering. However, for the mock predictions we will always plot the error computed from 
the scatter between the mocks. This is not only true for the clustering measurement, but also for the errors on the parameters of our powerlaw fits. When we plot the mean correlation 
functions of our 26 mock realisations, we do not divide  our error estimates by $\sqrt{26}$, such that the errors reflect the scatter expected on an individual realisation of the 
GAMA survey. Again recall these errors are underestimates in the two higher redshift slices due to the overlap between lightcones of different realisations in the simulation. 

When assessing the goodness-of-fit of the mock predictions to the data, or our data to literature data, we will use the covariance matrix. For the comparisonsof our GAMA observations to the literature we use the covariance matrix from the jack-knife resampling. As we compare to the SDSS measurements of \citet{zehavi11}, we can disregard the contribution of the errors on the literature measurements to the goodness-of-fit as they come from a much larger volume than our GAMA samples. When comparing the mocks to the data, we use the covariance matrix computed from the 26 mock realisations. This tests the hypothesis that the data is a realisation of the mock. As with the SDSS measurements, the error on the mean measurements from the 26 mocks are small compared the real data uncertainties and are therefore disregarded.

The computation of the inverse of the covariance matrix, $C_{ij}^{-1}$, can cause problems as our estimates of the covariance, from a limited number of resamplings 
or realisations, can be noisy. Our method to correct for this follows \citet{gaztanaga05} and \citet{marin13}. We first compute the correlation matrix,
\begin{equation}
 \tilde{C}_{ij} = \frac{C_{ij}}{\sqrt{\sigma_{i}\sigma_{j}}}
\end{equation}
where $\sigma_{i}$ is the standard deviation of the $i^{th}$ 2PCF measurement and $i$ and $j$ are indices running over all of the 2PCF measurements. We 
then carry out a singular value decomposition (SVD) of it, yielding $\mathbf{\tilde{C}}=\mathbf{U} \mathbf{\tilde{C}_{\rm{SVD}}} \mathbf{U^{T}}$, where 
$\mathbf{U}$ and $\mathbf{U^{T}}$ are rotation matrices and $\mathbf{\tilde{C}_{\rm{SVD}}}$ is a diagonal matrix with elements $\lambda_{ij}^2 \delta_{ij}$. The 
rotation matrix acts to transform the data points into a coordinate system where they are no longer correlated. The basis of this new coordinate
system are the eigenvectors of the covariance matrix, given by the columns of the matrix $\mathbf{U}$, which have $\lambda_{ij}^2 \delta_{ij}$ as
their eigenvalues. As explained in \citet{gaztanaga05} the eigenmodes of the data expressed in this basis should be uncorrelated, Gaussian
distributed and given by
\begin{equation}
\hat{W}(i) = \Sigma_{j} U_{ji}\frac{W(j)}{\sigma_{j}} 
\end{equation}
where $W(j)$ are elements of a vector of the 2PCF measurements, and $\hat{W}(i)$ are the new eigenmodes. As we have a noisy estimate of the covariance matrix, from a 
limited number of jack-knife samples or mock realisations, some of the eigenvectors will be poorly estimated and using them would bias the estimated
$\chi^2$. We would expect the modes which contribute least to the the variance to be most likely to suffer from this problem; these modes have small 
eigenvalues.

We experimented with applying the \citet{gaztanaga05} cut of $\lambda_{ij}>\sqrt{2/N_{\mathrm{mock,jk}}}$ where $N_{\mathrm{mock,jk}}$ is either the number of mock realisations for
the $\chi^2$ tests of the model ($N_{\mathrm{mock,jk}}=26$), or the number of jack-knife resamplings for the $\chi^2$ tests of whether our results agree with
\citet{zehavi11} ($N_{\mathrm{mock,jk}}=9$). However, we found that this could, counter intuitively, lead to noisier data sets or data sets with fewer reliable measurements 
having more accepted modes. We therefore adopted the approach of only taking the 4 largest modes, which for our data is slightly more conservative than applying the \citet{gaztanaga05} cuts.
 
To invert $\mathbf{\tilde{C}_{\rm{SVD}}}$ we simply take $1/\lambda_{ij}$ for each element, this is correct as the matrix is diagonal. We
set $1/\lambda_{ij}=0$ for eigenvalues failing the cuts. Setting values to zero in this way means some possible degrees of 
freedom are removed. We compute $\chi^{2}$ using the deviations divided by their associated errors, i.e.
\begin{equation}
\chi^{2} = \Sigma_{i,j} \frac{(W(i)_{\mathrm{data}} - W(i)_{\mathrm{ref}})}{\sigma_{i}} \tilde{C}_{\mathrm{SVD},ij}^{-1}\frac{(W(j)_{\mathrm{data}} - W(j)_{\mathrm{ref}})}{\sigma_{j}}.
\end{equation}
Here the `data' measurements are the measured GAMA 2PCF and the `ref' measurements are either literature data or the 2PCF prediction from the combined
mock catalogues. As our $\chi^2$ values are still likely to be somewhat inaccurate, we will only indicate whether the cumulative $\chi^2$ distribution
for four degrees of freedom suggests a probability of a certain measurement is less than $2\%$, demonstrating it shows highly statistically significant differences. 

\section{Results}\label{sec:resultsGAMA}
\subsection{Comparison to literature results}
In this section we compare our measurements to those of \citet{zehavi11}, which come from the SDSS. This comparison acts as a test that our methods for 
$k$-correcting galaxies, producing randoms and measuring clustering give reasonable results. We can also use the large area of the SDSS to gauge if the 
GAMA volume is particularly under-dense or over-dense.
\begin{figure*}
 \includegraphics[width=1.0\textwidth]{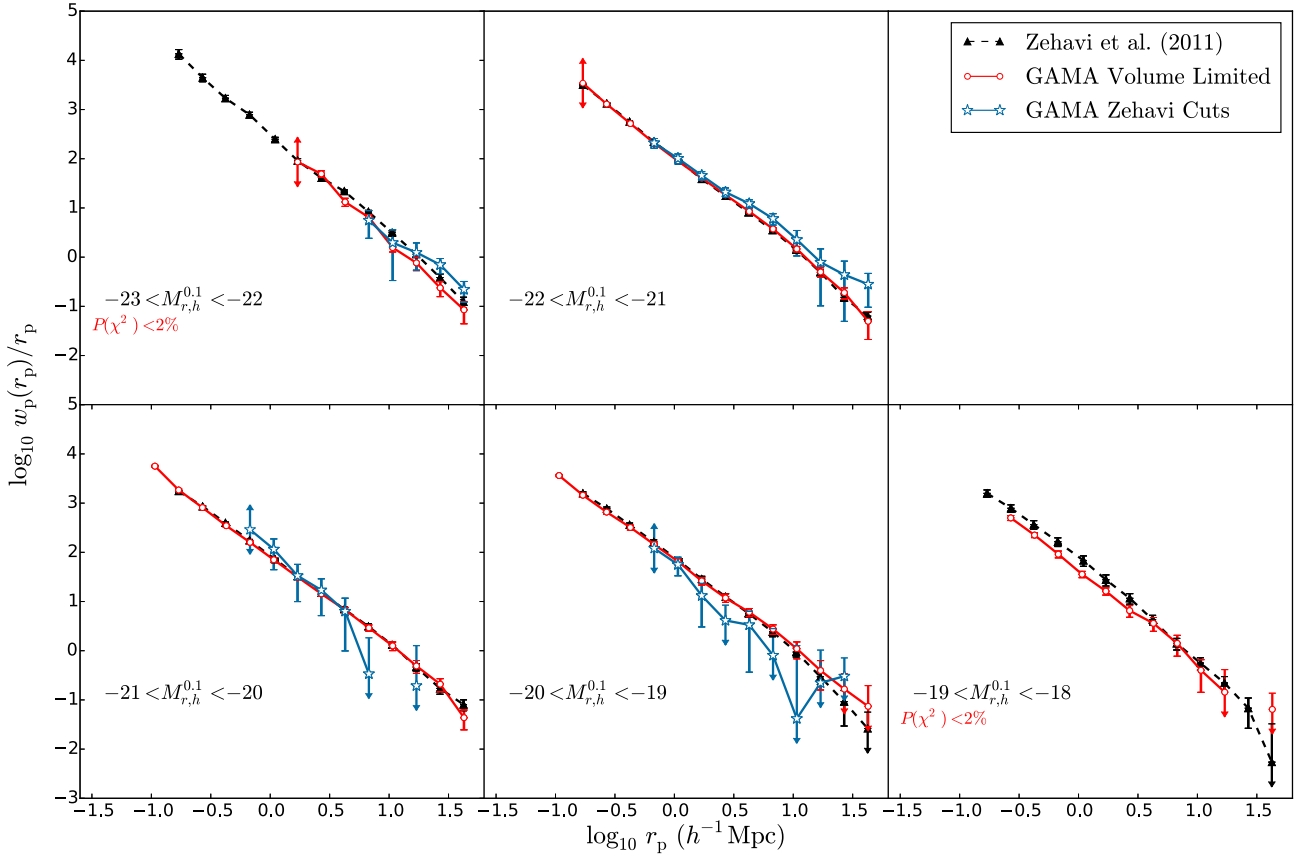}
 \caption{Clustering measurements from Zehavi et al. (2011) (black), along with our measurement of clustering for the same magnitude and redshift cuts 
(blue) and for our, deeper volume-limited samples (red). Error bars are from jack-knife resamplings of the data. As GAMA has a smaller area than SDSS, applying 
the same magnitude and redshift cuts as Zehavi et al. (2011) resulted in the fainter samples being too small to measure clustering. Where the
probability of our data being a realisation of the Zehavi et al. (2011) data falls to less than 2\%, this is indicated on the panel with the label $P(\chi^2)<2\%$, colour
coded by the sample tested. The measurements using the same redshift limits as Zehavi et al (2011) all pass this criteria.}\label{fig:zehavi}
\end{figure*}

In Fig.~\ref{fig:zehavi} we show our measurements (blue) against those of \citet{zehavi11} (black), both for the same redshift cuts. We also show, in red, our measurements 
for deeper volume limited samples using the GAMA apparent magnitude limit in Fig.~\ref{fig:zehavi}. 

When the probability of the hypothesis that these measurements agree with \citet{zehavi11} falls below 2\% it is indicated on the panel. As a simplification we do not use the \citet{zehavi11} error 
estimates in this calculation, as our errors are the dominant source of uncertainty. The samples with the SDSS redshift limits show good agreement with the SDSS measurements, suggesting the GAMA volume 
is not particularly unusual in the redshift ranges probed by these samples. 

The majority of our deeper, GAMA volume-limited samples agree with the \citet{zehavi11} measurements. The brightest however shows a significant variation compared to the lower redshift \citet{zehavi11} measurements. 
Note, however, that the jack-knife covariance matrices may not be accurate for a sample with such a small size and large Poisson errors. The faintest sample also appears 
to have a highly statistically significant difference compared to the \citet{zehavi11} measurement. Specifically, the clustering signal has a lower amplitude. Note, however, the volume of this
sample is even smaller than our low-$z$ slice, for which Fig.~\ref{fig:mockComp} demonstrates the jack-knife errors can be underestimates by a factor of $\sim 3$. In support of 
this being just being due to sample variance \citet{driver11} find that the GAMA survey is 15\% under-dense compared to SDSS DR7 up to $z=0.1$, and the median redshift of this sample is $z=0.09$.

\subsection{Stellar mass dependent clustering}\label{sec:massResults}
\subsubsection{The full shape of the correlation functions}
\begin{figure*}
 \includegraphics[width=1.0\textwidth]{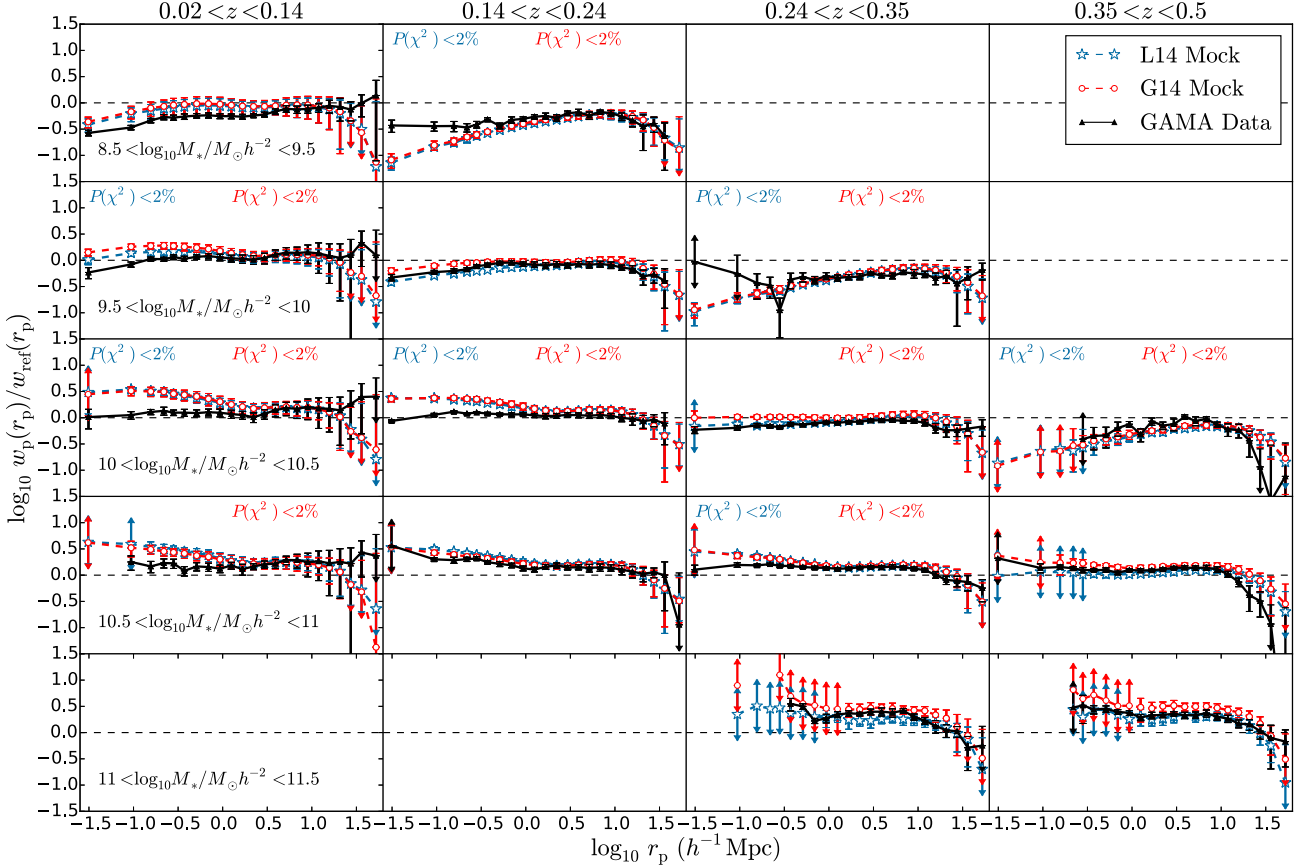}
 \caption{The projected two-point correlation function of real GAMA (black), G14 model (red) and L14 model (blue) galaxies as a function of redshift (different columns) 
  and stellar mass (different rows). The measurements are divided by the reference power law defined in Section~\ref{sec:proClus}. Error bars 
are from jack-knife resamplings for the real GAMA data and the scatter between the mock realisations for the mocks. Marked with $P<2\%$ are samples for which the hypothesis that 
the data is a realisation of the model has less than a $2\%$ probability, colour coded according to the model. Despite clustering measurements not being used when creating the models, they reproduce the same trends with mass and redshift as
 the data. Some samples, however, do show significant differences, particularly on small scales.}\label{fig:massCl}
\end{figure*}

We will now begin to study clustering as a function of a quantity that requires less modelling in the mocks but more modelling in the data: stellar mass. In 
Fig.~\ref{fig:massCl} we show the clustering of galaxies as a function of mass, divided by our reference power law. Dividing by 
a reference power law in this way means our measurements show the square of the galaxy bias, relative to the fiducial power law. Note that not all of the 
samples are volume limited, so one should be careful to use Table~\ref{table:gamaSamps1} to characterise the typical masses, luminosities and colours of the 
samples. Samples that are volume limited are indicated by a star in Table~\ref{table:gamaSamps1}. Samples which are more effected by the magnitude selection 
are expected to have less clustering than volume-limited samples \citep{meneux08}, and are expected to be bluer due to our colour-dependent $k$-corrections. 
This can be seen in Table~\ref{table:gamaSamps1}, where the higher redshift, more magnitude limited, samples of a particular magnitude or mass range tend to 
have bluer median colours. The mocks have been constructed to have the same selection function, so it is fair to compare directly.

In all the redshift slices we observe more clustering in the more massive galaxy samples. This is only seen in Fig.~\ref{fig:massCl} with volume-limited samples in the lowest redshift slice. However, in the two highest redshift slices the median colours and magnitudes of the second most massive sample, which is magnitude limited, are fairly similar to those of the low redshift volume-limited sample (Table~\ref{table:gamaSamps1}). This similarity suggests that the large change in clustering amplitude seen between the $10.5<\log_{10}M_{*}/\mathrm{\mdot}h^{-2}<11.0$ sample and the $11.0<\log_{10}M_{*}/\mathrm{\mdot}h^{-2}<11.5$ sample is real, even though the less massive of these samples is magnitude-limited. In the redshift slices we also see the small-scale clustering (around  $<3\,h^{-1}\mathrm{Mpc}$), relative to the large-scale clustering, increases with mass. This could be interpreted as an increasing fraction of satellite galaxies in the higher mass samples, or it could be a result of fainter, redder satellite galaxies being lost in the more magnitude-limited samples.

The mock predictions for the two models are fairly similar to one another. This similarity suggests the adoption of the
\citet{jiang08, jiang14} approach to computing satellite merger times has only a small effect on the clustering. The general trend of more massive 
galaxies being more clustered is reproduced. In a $\Lambda$CDM cosmology this implies in both the model and the data more massive galaxies reside in 
more massive dark matter haloes. As discussed, some of this effect may also be due to using magnitude-limited rather than
volume-limited samples. While the amplitude of the clustering in the models is generally an acceptable match to the data, the small-scale 
clustering is often incorrectly predicted. 

For the two slices below $z=0.24$, galaxy samples in the $10.0<\log_{10}M_{*}/\mathrm{\mdot}h^{-2}<10.5$ mass range appear, 
by eye, to have too high small-scale clustering. This discrepancy is borne out statistically, with values less than $2\%$ probability of this data
being a realisation of the model. For the G14 model this over-prediction of small scale clustering persists to higher redshift. The mock predictions
for the $10.0<\log_{10}M_{*}/\mathrm{\mdot}h^{-2}<10.5$ sample at $0.35<z<0.5$ also show a significant discrepancy, which does not
appear to be restricted to small scales. However recall that the covariance matrices derived from the mocks are underestimates of 
the true error in the high redshift slice, due to the oversampling of the N-body simulation. 

Additionally, the $8.5<\log_{10}M_{*}/\mathrm{\mdot}h^{-2}<9.5$ sample in the $0.14<z<0.24$ slice and the $9.5<\log_{10}M_{*}/\mathrm{\mdot}h^{-2}<10$ in 
the $0.24<z<0.35$ slice has a clustering signal on scales less than a few Mpcs that is significantly too small compared to the GAMA data. This is true even for the latter of those samples, where the data error bars might suggest the model is a good fit. This is because the error bars are much smaller in the mock predictions than in the real data, as the mock number densities are much larger for this sample (recall only the luminosity functions were forced to match). 

\begin{figure*}
 \includegraphics[width=1.0\textwidth]{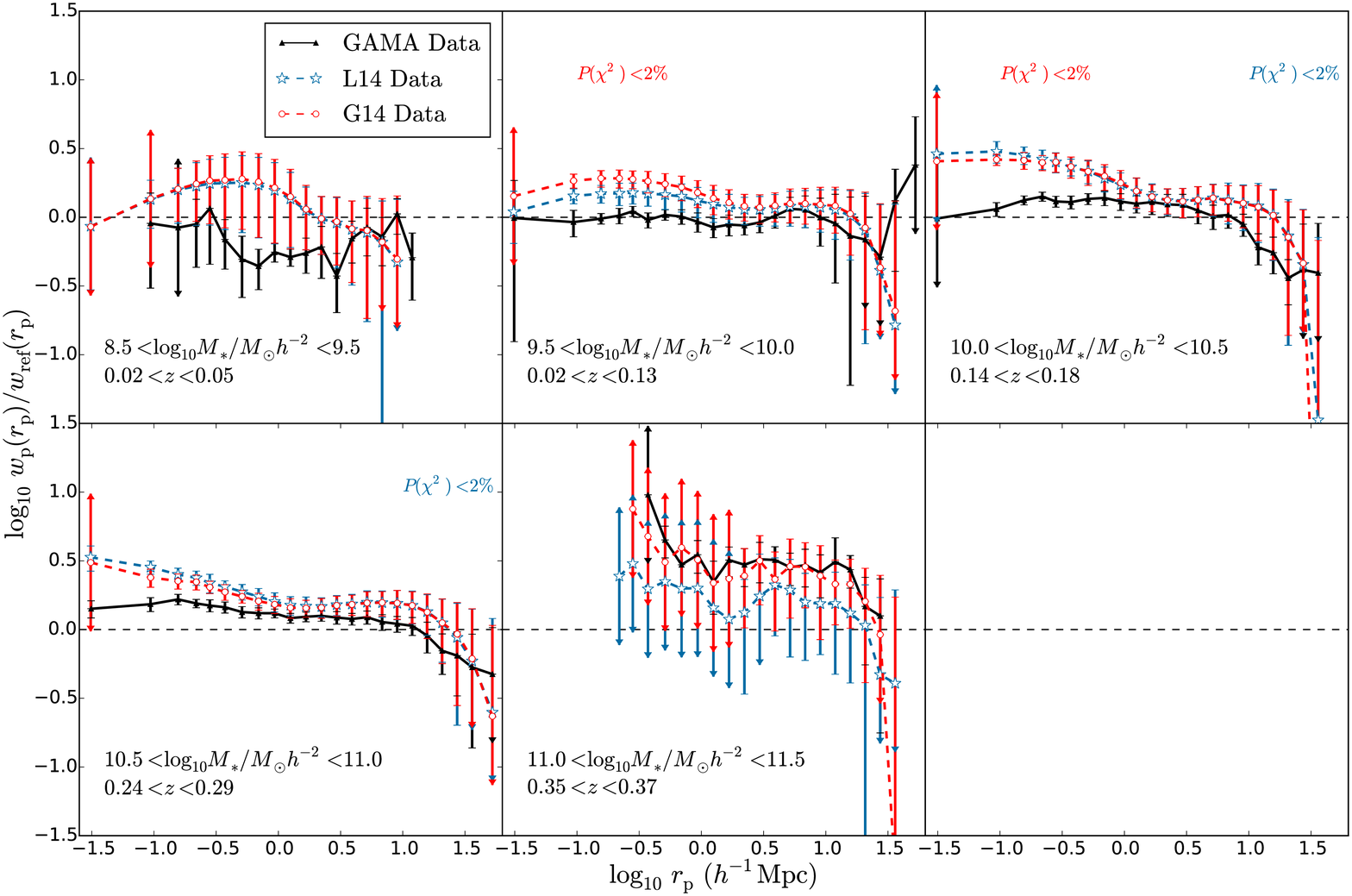}
 \caption{The projected two-point correlation function of real GAMA (black), G14 model (red) and L14 model (blue) galaxies as a function of stellar mass for 
 different approximately volume-limited redshift slices. The real GAMA samples consist of at least 98\% volume-limited galaxies, the mocks 89\%. The 
 measurements are divided by the reference power law defined in Section~\ref{sec:proClus}. Error bars are from jack-knife resamplings for the real GAMA data and the scatter between the mock realisations for the mocks. Marked with $P<2\%$ are samples for which the hypothesis that 
 the data is a realisation of the model has less than a $2\%$ probability, colour coded according to the model.}\label{fig:massClvlim}
\end{figure*}

In Fig.~\ref{fig:massClvlim} we plot samples which have had their upper redshift limits adjusted in order to make them volume limited. In general 
they imply the same conclusions as we drew from the flux-limited samples. We again see that the more massive galaxy samples are more clustered. Several of the 
mock samples have too much clustering on small scales. Seeing this effect in the volume-limited samples means it is not related to the modeling of the selecton function.


\subsubsection{Results from power law fits}\label{sec:powerlawmass}
\begin{figure*}
\begin{center}
 \includegraphics[width=1.0\textwidth]{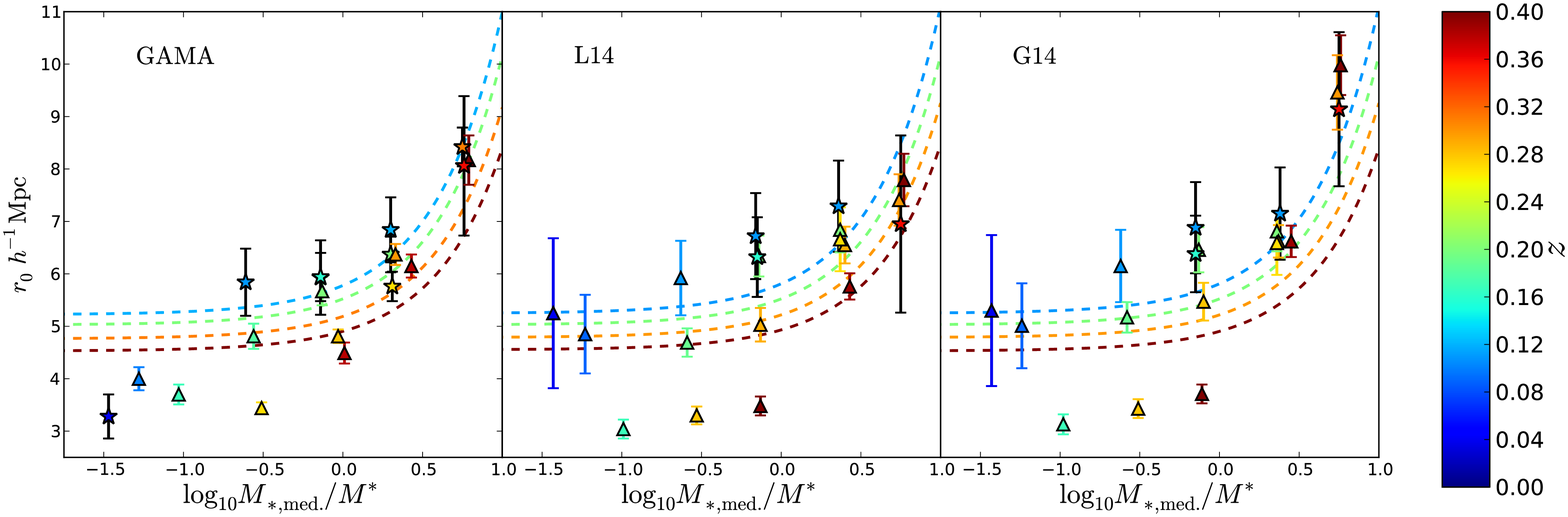}
 \caption{The $r_{0}$ of our power-law fits to the range $0.2\,\mathrm{Mpc}h^{-1}<r_{p}<9.0\, h^{-1} \mathrm{Mpc}$ as a function of the median stellar mass 
   of the sample divided by the characteristic mass of the stellar mass function (at $z=0$). The errors come from fitting to the multiple jack-knife 
   regions, or from the scatter from fitting to multiple mocks. Different panels give the GAMA measurements, the L14 model and the G14 model (left to right). Points have 
   been colour coded by median redshift, with samples defined by the same redshift cuts connected by lines. The dashed coloured lines 
   gives the bias fitting formula of \citet{li06}, converted from bias to $r_{0}$ assuming a power law correlation function with $\gamma=1.8$ and 
   evolved to different redshifts using the passive evolution model of \citet{fry96}, more 
   details are given in the text. More massive galaxies are more clustered, a trend qualitatively 
   reproduced by the models. Volume-limited samples are marked with a star, magnitude-limited samples are
   marked with a triangle. Note the magnitude-limited samples are likely to give lower values of $r_{0}$ than the corresponding volume-limited sample. As such the strong trend with redshift for the low 
   mass samples should not be interpreted as clear evidence of real redshift evolution. This selection effect is accounted for when constructing the model catalogues, 
   so the models and data can be fairly compared.}\label{fig:massro}
\end{center}
\end{figure*}

To further investigate the redshift and mass evolution of the clustering, we show in Fig.~\ref{fig:massro} the $r_{0}$ values from the power-law fits to the measurements, as a 
function of the ratio of the median sample mass to the characteristic mass of the stellar mass function at $z=0.0$, $10^{10.35} \rm{\mdot} /h^{2}$ \citep{baldry12}. Samples consisting of 95\% or more volume-limited galaxies are indicated by a star, magnitude limited samples are indicated by a triangle. We 
see very similar dependence of clustering on mass in both the GAMA data and the models. Also, in the magnitude-limited samples of the data and the models the increase in clustering strength as a function of stellar mass is faster for the higher redshift samples. This effect is 
also seen in the mock catalogues. This can also be viewed as less massive samples evolving faster with redshift than more massive ones. However,
note this effect is not seen when considering the volume-limited samples, which only show weak evidence of redshift evolution.

In order to further explore possible redshift evolution, in Fig.~\ref{fig:massro} we plot the bias versus mass relation of \citet{li06} as dashed lines, 
colour-coded according to redshift. We convert this relation into $r_{0}$ values assuming a power-law correlation function, and calculating the $r_{0}$ of 
dark matter using the bias and $r_{0}$ of the galaxy sample containing $M^*$, i.e. $r_{0,\mathrm{dm}} = r_{0}(M^*)b(M^*)^{-2/\gamma}$. We can then use the 
relation $r_{0}(M) = (b(M))^{2/\gamma}r_{0,\mathrm{dm}}$. We use our own power law fit to the published \citet{li06} sample containing $ M^*$ for 
$r_{0}(M^*)$. A corresponding bias, $b(M^*)$, is taken directly from the  \citet{li06} fitting formula. The dashed lines at different redshifts are calculated using the passive evolution model of \citet{fry96} to evolve the bias values, and using the growth factor $D(z)$ to evolve the dark matter $r_{0}$ as $r_{0,\mathrm{dm}}(z) = r_{0,\mathrm{dm}}(z')(D(z)/D(z'))^{2/\gamma}$. Such a model gives the expected evolution of $r_{0}$ for a model where galaxies
formed in some density field before moving along trajectories at a velocity defined by their local gravitational potential. The growth factor was calculated 
for our cosmology from the approximate formula of \citet{carroll92}. We assume $\gamma=1.8$ for the power laws, whilst in reality our value of $\gamma$ is allowed to vary in the fits. However, we show in Table~\ref{table:fits} that using a fixed $\gamma=1.8$ only has a small affect on our $r_{0}$ values. 

 One clear observation is that in both the volume-limited and magnitude-limited lowest mass sample the measured amplitude is much lower than the 
\citet{li06} relation. As previously mentioned, GAMA is unusually underdense at low redshifts, and the jack-knife error bars can considerably 
underestimate the true uncertainty at low redshift. The very low amplitude in the least massive samples is therefore likely sample variance.

It appears at lower masses the dependence on $r_{0}$ with mass is stronger in our data than the fitting formulae. Some of this may be sample variance, but note this strong trend is not seen when 
only considering volume-limited samples. Indeed the volume-limited samples follow the expected dependence of
clustering amplitude with mass.  Another likely selection effect is that there is more redshift evolution in the magnitude-limited samples than expected from 
the passive evolution model. In Fig.~\ref{fig:massro} we help to mitigate the effect of the GAMA $r$-band magnitude selection on our results by plotting 
$r_{0}$ against the median mass of the sample. Nonetheless, higher redshift samples will be biased toward different galaxies. As mentioned, more 
magnitude-limited samples will have lower clustering than volume-limited samples \citep[e.g.][]{meneux08}. As such the redshift evolution in clustering 
amplitude observed in the less massive galaxies may be down to these observational effects. Tests we conducted comparing clustering measurements from 
$r<19.8$ and $r<21$ versions of the mock catalogues support this idea (these samples have a much larger fraction of magnitude-limited galaxies 
than the samples where we found no significant differences between the original and deeper catalogues).

At higher masses our measurements are not precise enough to determine if the galaxy clustering is evolving passively or not. It is likely that mass samples in these data can be used to probe redshift evolution, but this may be better done using more sophisticated modeling such as HOD fitting. We leave this to the HOD fitting of Palamara et al (in prep), while we focus on comparison to the models which have selection effects included via the use of lightcone.

Returning to the model comparisons then, we see the $r_{0}$ values of each model are very similar to one another, except in the high redshift slice where the L14 model has stronger clustering than the G14 model. The trends seen in the data are also reproduced by both models. In general the increase of $r_{0}$ with magnitude is slightly steeper in the model than the data. The steep increase of clustering with mass, at low masses, is expected to be related to the GAMA 
selection function. As this trend is qualitatively reproduced by the models, the models have some success in assigning the correct luminosities or colours to the sample galaxies. 

\subsubsection{Summary of model comparisons for mass samples}
To summarize, the most obvious problems with the model predictions as a function of stellar mass are in the one-halo term regime, this points towards the physics of 
satellite galaxies being a weakness in the model. As mentioned, in \citet{campbell15} instead of using the true model masses, they estimate the model masses from 
the predicted broad-band photometry. They find that this does affect the clustering as a function of mass. The effect is much stronger for the L14 model than the G14 model, 
and brings the small scale clustering into better agreement with the data.  They still find that even by estimating the masses in this way, the models predict too much 
small-scale clustering. Also note that the typically redder and fainter satellite galaxies are more sensitive to the $r$-band apparent magnitude cut (i.e. selection effects), so it may also 
be related to the model assigning wrong colours or luminosities to satellite galaxies. This is also a possible explanation for the steeper increase in $r_{0}$ with mass compared to real data.  

\subsection{Luminosity dependent clustering}
\subsubsection{The full shape of the correlation functions}
\begin{figure*}
 \includegraphics[width=1.0\textwidth]{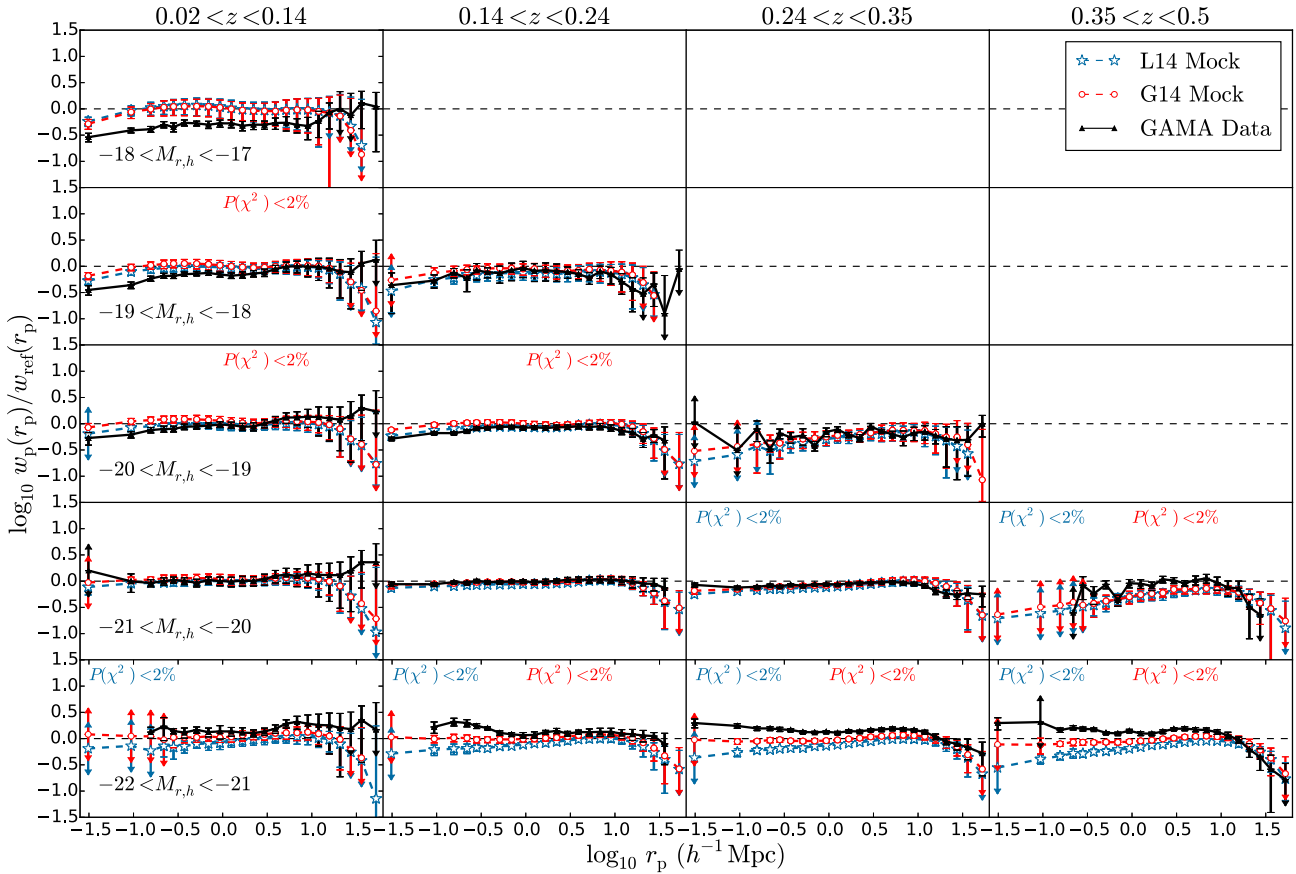}
 \caption{The projected two-point correlation function of real (black) and G14 model (red) and L14 model (blue) galaxies as a function of redshift (
  different columns) and luminosity (different rows). The measurements are divided by the reference power law defined in Section~\ref{sec:proClus}. Error 
  bars are from jack-knife resamplings for the GAMA data, and from the scatter between realisations in the mock. Marked with $P<2\%$ are samples for which 
  the hypothesis that the data is a realisation of the model has less than a $2\%$ probability, colour coded according to the model. The clustering of faint 
  galaxies is successfully reproduced in the L14 model for the majority of samples. However, the models do not show as much clustering evolution with 
  luminosity as the data, and are particularly discrepant for the brightest sample and on small scales.}\label{fig:lumCl}
\end{figure*}

\begin{figure*}
 \includegraphics[width=1.0\textwidth]{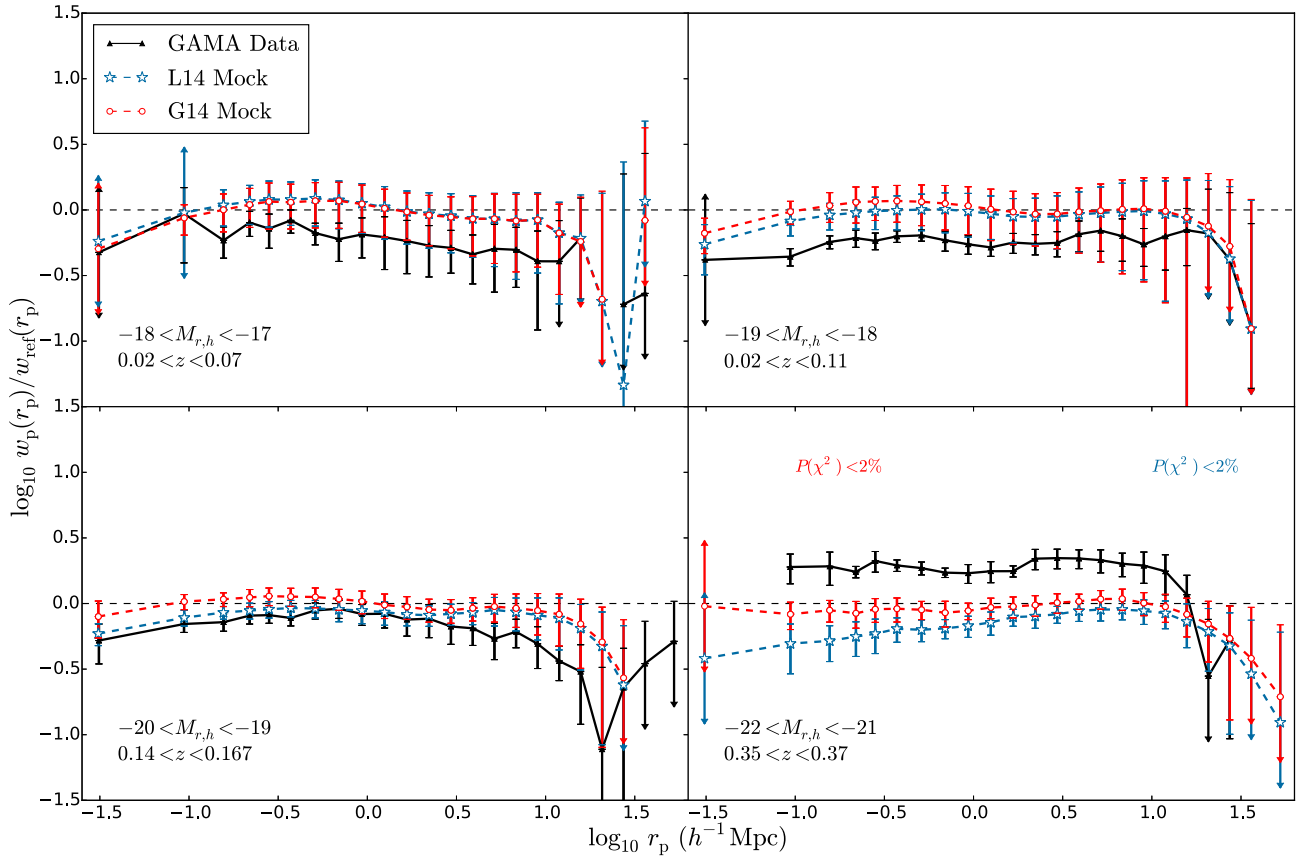}
 \caption{The projected two-point correlation function of real (black) and G14 model (red) and L14 model (blue) galaxies for different
  volume-limited luminosity samples. The measurements are divided by the reference power law defined in Section~\ref{sec:proClus}. Error 
  bars are from jack-knife resamplings for the GAMA data, and from the scatter between realisations in the mock. Marked with $P<2\%$ are samples for which 
  the hypothesis that the data is a realisation of the model has less than a $2\%$ probability, colour coded according to the model.}\label{fig:lumClvlim}
\end{figure*}

In Fig.~\ref{fig:lumCl} we show the clustering of galaxies as a function of luminosity and redshift, in Fig.~\ref{fig:lumClvlim} we show the 
corresponding volume-limited samples. For all of the redshift intervals we notice 
segregation between the faint and bright samples, with brighter galaxies being more clustered. This trend is not seen in the mocks,
indeed the clustering of the L14 mock decreases slightly between $-21.0<M_{r,h}<-20.0$ and $-22.0<M_{r,h}<-21.0$. Because of this lack of dependence on
luminosity, the mocks brighter than our sample $-21.0<M_{r,h}<-20.0$ (which contains  $M^{*}_{r,h}$) all disagree very significantly with the data. For these
bright galaxies the G14 model is closer to the data than the L14 model, but still very discrepant. 

For bins fainter than  $M^{*}_{r,h}$, we see good agreement for many of the L14 model samples, in particular for samples fainter than $M_{r,h}=-20$. In 
contrast, some of the G14 model predictions for fainter luminosity samples show too much clustering on small scales (less than $\sim 3\,h^{-1}\mathrm{Mpc}$). At $0.24<z<0.35$ and $-21.0<M_{r,h}<-20.0$ the L14 model predicts too little small-scale clustering. Additionally, the most luminous mock samples under 
predict the clustering, particularly on small scales ($<3h^{-1}\,$Mpc) at $z>0.14$. Apart from this most luminous 
sample, the small scale clustering predictions of the mocks are more successful here than in the mass samples.
The volume-limited samples in Fig.~\ref{fig:lumClvlim} all show reasonable agreement, given the errors, with the model except the most luminous
sample. In this sample, as with the magnitude-limited sample around this redshift, the models underpredict the clustering.
\begin{figure*}
\begin{center}
 \includegraphics[width=1.0\textwidth]{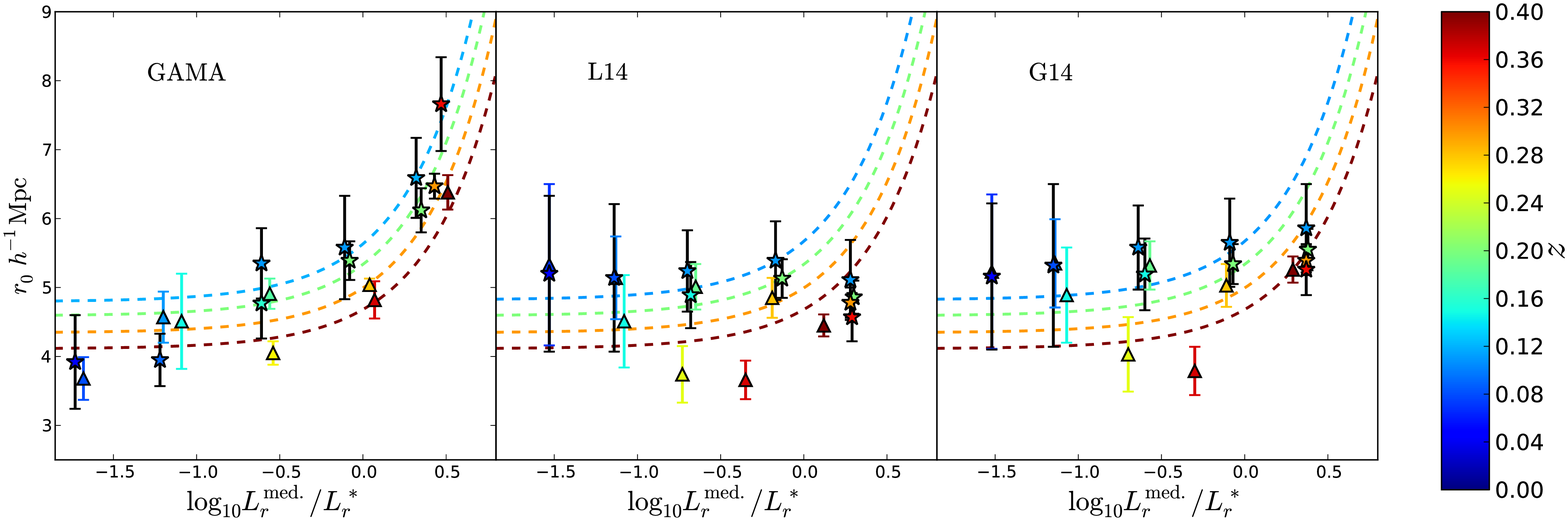}
 \caption{The $r_{0}$ of our power law fits to the range $0.2<r_{p}<9.0\, h^{-1}\mathrm{Mpc}$ as a function of the median luminosity of the sample, divided by the 
   characteristic luminosity of the data at $z=0$. Pure luminosity evolution has been accounted for with Q=1.45. The error bars give the error from fitting to the multiple 
   jack-knife regions. Different panels give the GAMA measurements, the G14 model and the L14 model (left to right). Points have been colour coded by median redshift, with 
   samples defined by the same redshift cuts connected by lines. Volume-limited samples are plotted as stars, magnitude-limited samples as triangles. The dashed coloured lines 
   gives the bias fitting formula of \citet{zehavi11} with $\sigma_{0.8}=0.8$, converted from bias to $r_{0}$ assuming a power law correlation function with $\gamma=1.8$ and 
   evolved to different redshifts using the passive evolution model of \citet{fry96}, more 
   details are given in the text. The \citet{zehavi11} fitting formula is only derived from the range $0.16<L_{r}/L_{r}^{*}<6.3$ \citep{zehavi11}, but the disagreement at 
   low luminosity is more likely to be down to sample variance underestimated in the jack-knife errors. Brighter galaxies are more clustered in the data, a trend not 
   reproduced in the models. Note the magnitude-limited samples are likely to be underestimates of the true $r_{0}$. As such the trends with redshift for the low 
   luminosity samples should be interpreted with caution (see the text). In the highest luminosity samples, three volume-limited samples demonstrate that $r_{0}$ is evolving
   slower than the passive evolution model (i.e. the bias is evolving faster).}\label{fig:lumro}
\end{center}
\end{figure*}
\subsubsection{Results from power law fits}\label{sec:powerlawlum}
In Fig.~\ref{fig:lumro} we plot the $r_{0}$ of our luminosity samples, as a function of the ratio of the median sample luminosity to $M^{*}_{r,h}=-20.6$ 
\citep{loveday15}. Taking this ratio is useful as it removes a dependence on how magnitudes are converted to luminosities. We see less variation in the clustering 
properties over this range than we did with mass. Volume-limited samples are plotted with a star, magnitude-limited samples are plotted with a triangle.  In 
Fig.~\ref{fig:massro} we plot the \citet{zehavi11} bias versus luminosity fitting formula as a dashed line, colour coded according to what redshift the 
relation has been passively evolved. Following Section~\ref{sec:powerlawmass}, to determine the $r_{0,\mathrm{dm}}$ needed to convert galaxy biases from the \citet{zehavi11} 
formula to $r_{0}$ values, we use the quoted $r_{0}$ value from the \citet{zehavi11} power law fit to their sample containing $L_{r}^*$ as a reference. 

Before interpreting these data, we must once more consider the selection effects of the magnitude-limited samples. For these samples the effect 
is smaller, as the luminosity range is limited in the sample. Nonetheless, because of the colour-dependent $k$-corrections the colours of the samples can 
change due to the magnitude selection. As before the selection effects may act to artificially enhance any redshift or luminosity trends, by making the 
clustering of fainter and more distant galaxies appear weaker than it would in a volume-limited sample. This certainly seems to be the case for
the most luminous sample in the highest redshift range, where the volume-limited sample has a much larger amplitude than the magnitude limited one at the
same redshift. Note the redshift range of the volume limited sample is very small, and as such it is likely to suffer from sample variance. The $r_{0}$ 
values of the other volume-limited samples seems to roughly agree with the magnitude-limited sample, but note this is for the subset of samples where producing
a volume-limited version was possible. It was not possible to construct volume-limited samples for samples with very high fractions of magnitude-limited 
members, where selection effects are likely to be largest.  
 
The data show an increase in clustering with luminosity in all of the redshift slices. This is seen in many of the volume-limited samples, and so cannot 
solely be down to selection effects. Unfortunately, as with the mass samples, this trend cannot be confirmed in the two highest redshift slices with 
volume-limited samples. Except for the lowest-luminosity sample our measurements from real GAMA data agree with the \citet{zehavi11} relation in the lowest 
redshift slice, which is close to the redshift at which the relation was measured. Our lowest luminosity measurement falls well below the \citet{zehavi11} relation. As with the lowest-mass sample, because we know GAMA is under-dense at $z<0.1$ \citep{driver11} (which is above the median redshift of the faintest 
sample), a far more likely explanation is sample variance unaccounted for by the jack-knife errors.

Except for the brightest sample, the $r_{0}$ values seem to roughly follow the expectations of passive evolution. Of course, however, one has to worry 
about the selection effects. Using the volume-limited samples alone only gives some weak evidence of redshift evolution in $r_{0}$ between the lowest and second
lowest redshift slices.

The highest luminosity sample shows some evidence of not following the passive evolution model. The volume-limited sample in particular
shows stronger than expected clustering from a passive evolution model. However, given the size of the uncertainties the measurements are only 1-2$\sigma$ away from the 
passive evolution model. A lack of evolution, compared to that expected from passive evolution, would suggest fast bias evolution. This would be consistent with many other observations of more massive and brighter galaxies showing little evolution in clustering with redshift \citep[e.g.][]{white07, brown08, coil08, meneux08}.

Moving back to our focus of model comparisons, the models show little dependence on clustering amplitude and luminosity. Both models predict a similar trend 
of $r_{0}$ versus luminosity. This lack of luminosity dependence is also in agreement with the \citet{zehavi11} bias fitting formula, up to the brightest 
sample where the L14 model $r_{0}$ values falls below the fitting formula. The models both predict much smaller $r_{0}$ values than the data in the highest luminosity bin, the models, unlike the data, also show evidence of evolution in the clustering with redshift in this sample.
 
\subsubsection{Summary of model comparisons for luminosity samples}
The models do a reasonable job at reproducing the amplitude of clustering as a function of luminosity for the samples fainter than $L_{r}^{*}$, and the 
L14 model is a good fit to many of the fainter samples. However, the models fail to reproduce the 
amplitude of our brightest sample. This under prediction particularly affects small-scale clustering. Also, recall \citet{kim09} found the dependence on luminosity was not correct for the \citet{bower06} model. Note that semi-analytic models have been shown to have a dependence on clustering with luminosity, but this evolution only starts to become apparent at the highest luminosity probed here \citep[see e.g.][]{norberg01}. Again, the disagreements are particularly large at small scales, suggesting the modelling of satellite galaxies needs to be improved in the models.

\subsection{Colour dependent clustering}
\begin{figure*}
\includegraphics[width=1.0\textwidth]{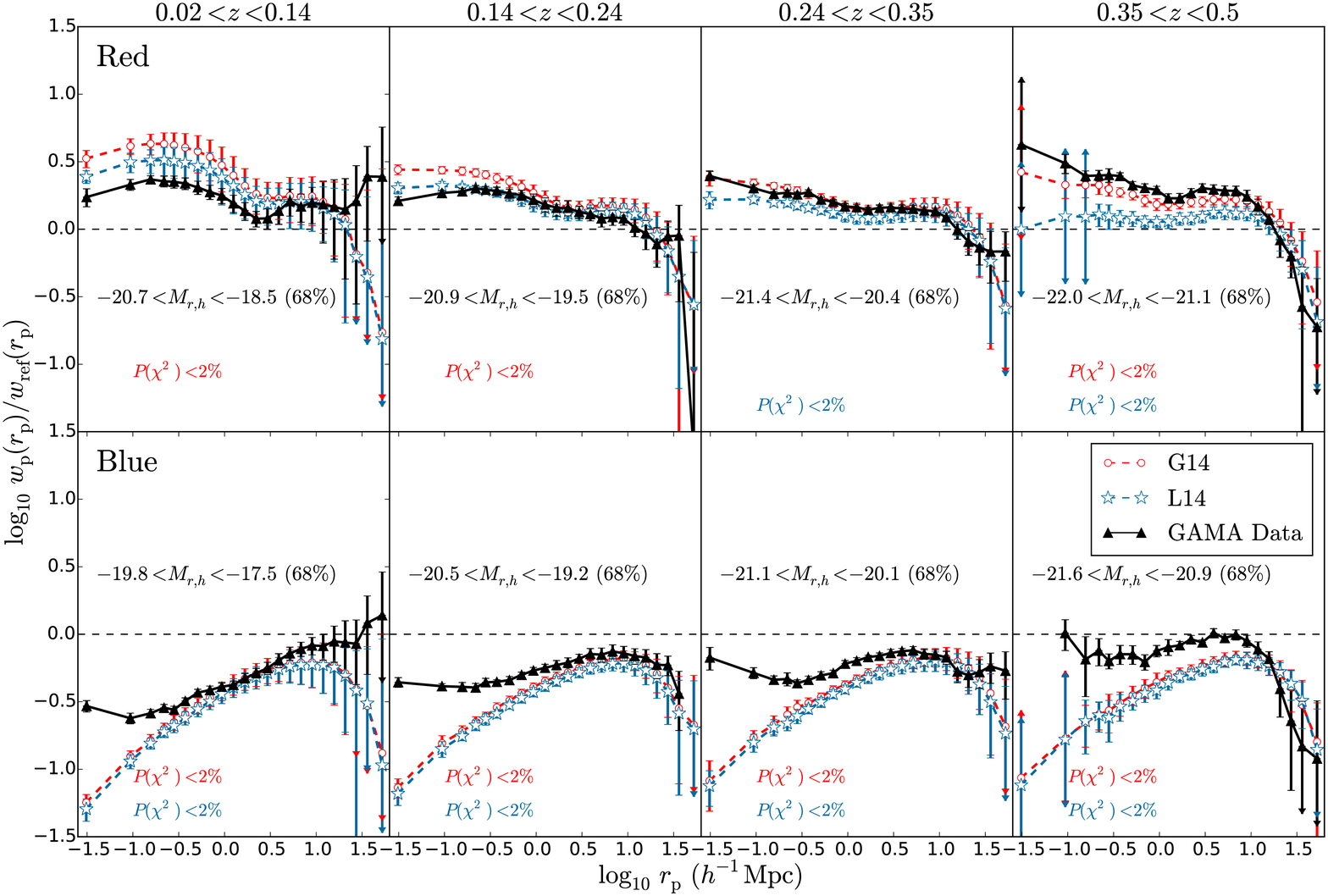}\caption{The projected correlation function of red galaxies (top) and blue galaxies (bottom) in 
 different redshift slices. Also shown is the clustering of mock galaxies for the L14 model (blue) and the G14 model (red). The results have been divided by the reference power 
 law defined in   Section~\ref{sec:proClus}. Marked with $P<2\%$ are samples for which the hypothesis that the data is a realisation of the model has less than a $2\%$ probability, 
 colour coded according to the model.}\label{fig:colourCl}
\end{figure*}

Fig.~\ref{fig:colourCl} shows the clustering of our samples of red and blue GAMA galaxies (in black). In all redshift intervals we probe, red 
galaxies are more clustered than blue galaxies. We also note that the red galaxy correlation functions have their strongest clustering, relative
to the reference line, at scales less than around $2$  $h^{-1}$ Mpc. This can be interpreted as red galaxies predominantly being in larger haloes with more 
satellite galaxies. The blue galaxy samples also show a relative increase in clustering on very small scales ($<0.3$~$h^{-1}$~Mpc), but note that their clustering
is still weaker than the reference power law.   

In each panel of Fig.~\ref{fig:colourCl}  we indicate the absolute magnitude range enclosing the central 68\% of the data. From  Fig.~\ref{fig:colourCl}, we can draw the following 
natural conclusions. Firstly, for a given redshift slice, red galaxies are typically brighter than blue ones. Additionally, the absolute magnitude range of the central 68-percentile 
reduces with redshift and gets brighter with redshift. Finally, at fixed redshift, the change in clustering between red and blue galaxies is significantly larger than can be explained 
by any luminosity (or stellar mass) dependence of clustering alone.
The models also show red galaxy samples have a larger amplitude and steeper 2PCFs than blue galaxies. At low redshifts ($z<0.24$) the L14 model agrees with the measurements whilst the
G14 model has too much small-scale clustering. In contrast the G14 model shows better agreement in the two high redshift slices, whilst the L14 model now has too little
small-scale clustering. 

The blue mock galaxy samples all have too low an amplitude, particularly at small scales. The size of the $\chi^{2}$ values can leave us in no doubt
that neither model successfully reproduces the clustering of blue galaxies. As the small scale measurements are particularly discrepant, it is possible 
that too few satellite galaxies are blue. As these samples are magnitude-limited, another possibility is the model predicts blue satellites that are
too faint. The predictions for blue galaxies being incorrect may not be too surprising, when one considers that the models 
predict a clear sequence of blue galaxies not seen in the data (Fig.~\ref{fig:gamaCMD}). The models also do not reproduce the increase in
clustering relative to the power law seen at very small scales ($<0.3~$$h^{-1}$~Mpc). This could relate to an increase in star formation
rate for close pairs of galaxies. This possibility will be explored in a future paper using star-formation-rate-selected samples 
of GAMA galaxies (Gunawardhana et al. in prep).

\section{Discussion and Conclusions}\label{sec:concGAMA}
We have studied the projected two-point correlation function of galaxies in GAMA. To do this we used a modified version of the \citet{cole11} approach to 
generate random catalogues; this method resulted in a set of random points with all of the properties of the real galaxies. The \citet{cole11} approach 
allows sample selection cuts to be applied to both the data and the random catalogue, allowing the measurement of galaxy clustering as a function of diverse galaxy 
properties. Our modification, the inclusion of a window to limit the difference between the initial and cloned redshift, has the potential to limit the effects of galaxy 
evolution on random catalogues. This method should be followed up in later work, to test how effective it is on a wider selection of 
galaxy samples (e.g. not just selected on optical photometry) and to optimize the size of window used. 

We compared volume-limited samples from SDSS of \citet{zehavi11} to volume-limited samples in GAMA, both with the redshift cuts of SDSS and with
redshift cuts appropriate for the deeper, GAMA magnitude limit. We find good agreement with SDSS for our samples using the SDSS redshift
limits. In our $-23.0<M_{r,h}^{0.1}<-22.0$ sample and in our faintest sample, $-19.0<M_{r,h}^{0.1}<-18.0$, both with the GAMA redshift limits, we find some 
disagreement between SDSS and GAMA. Given the expected uncertainties in our jack-knife covariance matrices, this could indicate the errors were 
underestimated. Our faintest sample, $-19.0<M_{r,h}^{0.1}<-18.0$, shows less clustering than the SDSS data. Particularly relevant for this sample is the \citet{driver11} 
observation that GAMA is under-dense at $z<0.1$, which would indeed lead to lower clustering.

We have observed that more luminous, more massive and redder galaxies are more strongly clustered, in redshift slices between $z=0$ and $z=0.5$. Though in
the two highest redshift slices the GAMA selection function complicates the interpretation of this trend. More massive and luminous 
galaxies being more clustered is in agreement with previous measurements that show these trends exist at lower and higher redshift 
ranges \citep[e.g.][]{zehavi11, christodoulou12, li12, torre13, marulli13, guo14}, and in a broad redshift bin encompassing all of our slices \citep{skibba14}. We also 
find that red galaxies have steeper correlation functions than blue galaxies in these redshift slices, again in agreement with clustering measured 
at higher \citep[e.g.][]{coil08, guo14} and lower \citep[e.g.][]{zehavi11} redshifts. 

We fit power laws to our measurements and see an evolution in the apparent clustering strength of galaxies with redshift for samples less massive than 
$10^{10.5}\mdot h^{-2}$. We also evidence of this occurring for galaxies less luminous than around $L^{*}$. The evolution is
in the direction of higher redshift galaxies being less clustered. Note however, many of our samples are not volume limited and so 
will be affected by selection effects that are very likely to mimic this evolution. The volume-limited samples alone do show some evidence of higher
redshift samples of the same luminosity or mass being less clustered, but this evidence is rather weak considering the size of the uncertainties. 

In our most luminous sample ($\sim 3L^{*}$), which is volume-limited in three of the redshift slices, we see little evolution in clustering amplitude between samples with median redshifts of $z \sim 0.1$ and $z \sim 0.4$. Given our uncertainties, there is some weak evidence that this is less evolution than that expected from a simple, passive evolution model.

A lack of evolution in the clustering amplitude of brighter and more massive galaxies has been observed by other authors, and taken as evidence of a fast 
evolution  of the bias of these objects \citep{white07, brown08, meneux08, meneux09}. This fast bias evolution has been connected to the 
merging and disruption of satellites \citep{white07, brown08}. 

Whilst the interpretation of these measurements is complicated by selection effects in the magnitude-limited samples, the strength and focus of this
work is model comparisons. We use a model that utilizes a lightcone, and assign $k$-corrections to the model galaxies in such a way
to mimic the selections function of the survey. We find that the L14 and G14 semi-analytic models successfully predict the trends of clustering as a 
function of stellar mass. Both models have similar predictions for this trend, suggesting it is insensitive to differences in the adopted physics. In detail there 
are places where the models disagree with the data, often in the regime of the one-halo term, which is too strong. \citet{campbell15} notes that
estimating model masses from the colours, rather than taking them from the model, helps alleviate this problem for the L14 model. 

We find the increase of clustering seen between our samples $-21.0<M_{r,h}<-20.0$ and $-22.0<M_{r,h}<-21.0$ is not reproduced by the models. This 
discrepancy is present in both models, but is a bigger problem for the L14 model. For the highest luminosity sample ($\sim 3 L_{r}^*$)
the models also predict too little small-scale clustering. The clustering of fainter galaxies is better modelled in
the data, with the L14 model being particularly successful at reproducing the clustering of the  $M_{r,h}>-20$ galaxy samples.

We have also shown that the trends of redshift evolution with mass and luminosity are qualitatively reproduced by the models 
(Fig.~\ref{fig:massro} and Fig.~\ref{fig:lumro}), in that the more massive and more luminous samples show less redshift
evolution. The redshift trends are most likely successfully reproduced thanks to the modelling of the GAMA apparent magnitude limit, 
showing the importance of applying observational effects when comparing galaxy formation models to data.
 
The clustering of red galaxies on large scales is successfully reproduced by the models. At $z>0.35$ the L14 model under predicts the 
clustering on small scales and at $z<0.35$ the G14 model over predicts clustering on small scales. Both models under-predict the clustering of blue 
galaxies, particularly on scales less than a few Mpc. The colour-magnitude diagram of the models is also very different to that of the data, the models
have a much more clearly bimodal colour distribution.

These models give different predictions from one another for samples defined by mass, photometric properties and colour. This highlights the importance of 
testing the clustering of models using a variety of properties for sample selection, or, as \citet{campbell15} suggests, inferring the mass of the model galaxies from their photometric properties. 

The models struggle to reproduce the clustering at small scales. Therefore, a suggested route to improvement is modifying the physics affecting
satellite galaxies. Several authors have experimented with adding processes which remove satellite galaxies, or modify their luminosity or colour 
\citep[e.g][]{font08, kim09, contreras13}. \citet{kim09} additionally found that processes affecting satellite galaxies can also change the 
dependence of clustering on luminosity. Modifying the satellite galaxy physics could also have an impact of the colour distribution and improve the 
clustering predictions of blue galaxies. Indeed, \citet{font08} found bluer satellites when applying a more realistic model of the stripping of 
gas from infalling satellite galaxies. Modifications to the satellite galaxy physics could also affect the over abundance of faint red galaxies in the models noted by 
\citet{tamsyn14}, or the excess of high-multiplicity groups of galaxies observed by \citet{robotham11}. Finally, changing the colours of satellite galaxies
can have knock-on effects for magnitude-limited samples, as model galaxies with red SEDs are $k$-corrected out of these samples at lower redshifts than blue galaxies. 
Such modifications should also be tested to see if they reproduce the lack of redshift evolution in the brightest samples where the physics affecting 
satellites galaxies is expected to play a role \citep{white07, brown08}.
 
In conclusion, the G14 and L14 models reproduce the trends of clustering with mass and successfully predict the amplitude of clustering for galaxies fainter than
$L_{r}^*$. They do, however, need improvement at small scales - in particular blue galaxies need stronger small-scale clustering. They also need to be improved in order to 
reproduce the clustering of galaxies slightly brighter than $L_{r}^*$, i.e. for our samples with median $r$-band magnitudes around $M_{r,h} \sim 21.3$. When these measurements are 
compared to models one must be careful to account for the selection effects in the magnitude-limited samples, for a realistic mock catalogue this could be as simple
as applying an $r<19.8$ apparent magnitude cut. An alternative is to choose the samples which are volume-limited. Indeed,
in a future GAMA paper we intend to use the methodology here to produce volume-limited magnitude and mass threshold selected samples and carry out a HOD 
analysis on them (Palamara in prep.). In order to facilitate further model tests against these measurements, our results will be made available online at 
publication\footnote[3]{http://icc.dur.ac.uk/data/}. 

\FloatBarrier
\section*{Acknowledgements}
DJF thanks Ariel Sanchez for useful discussions. 
PN acknowledges the support of the Royal Society through 
the award of a University Research Fellowship and the European 
Research Council through receipt of a Starting Grant (DEGAS-
259586). MJIB acknowledges financial support from the Australian 
Research Council (FT100100280). JL acknowledges support from the Science and Technology Facilities Council (grant number ST/I000976/1).
This work was supported by the Science and Technology Facilities Council [ST/L00075/1]. Data used in this paper will be 
available through the GAMA DB (http://www.gama-survey.org/) once the associated redshifts are publicly released.

GAMA is a joint European-Australian project based around a spectroscopic campaign
using the Anglo-Australian Telescope. The GAMA input
catalogue is based on data taken from the Sloan Digital
Sky Survey and the UKIRT Infrared Deep Sky Survey.
Complementary imaging of the GAMA regions is being obtained
by a number of independent survey programs including
GALEX MIS, VST KIDS, VISTA VIKING, WISE,
Herschel-ATLAS, GMRT and ASKAP providing UV to radio
coverage. GAMA is funded by the STFC (UK), the ARC
(Australia), the AAO, and the participating institutions.
The GAMA website is http://www.gama-survey.org/. 

This work used the DiRAC Data Centric system at Durham
University, operated by the Institute for Computational
Cosmology on behalf of the STFC DiRAC HPC Facility
(www.dirac.ac.uk). This equipment was funded by BIS National
E-infrastructure capital grant ST/K00042X/1, STFC
capital grant ST/H008519/1, and STFC DiRAC Operations
grant ST/K003267/1 and Durham University. DiRAC is
part of the National E-Infrastructure.

\bibliographystyle{mn2e}
\setlength{\bibhang}{2.0em}
\setlength\labelwidth{0.0em}
\bibliography{bibliography}
\label{lastpage}

\appendix

\FloatBarrier

\section{The effect of P, Q and windowed randoms on the correlation function}\label{sec:cTest}
\begin{figure}
\includegraphics[width=.45\textwidth]{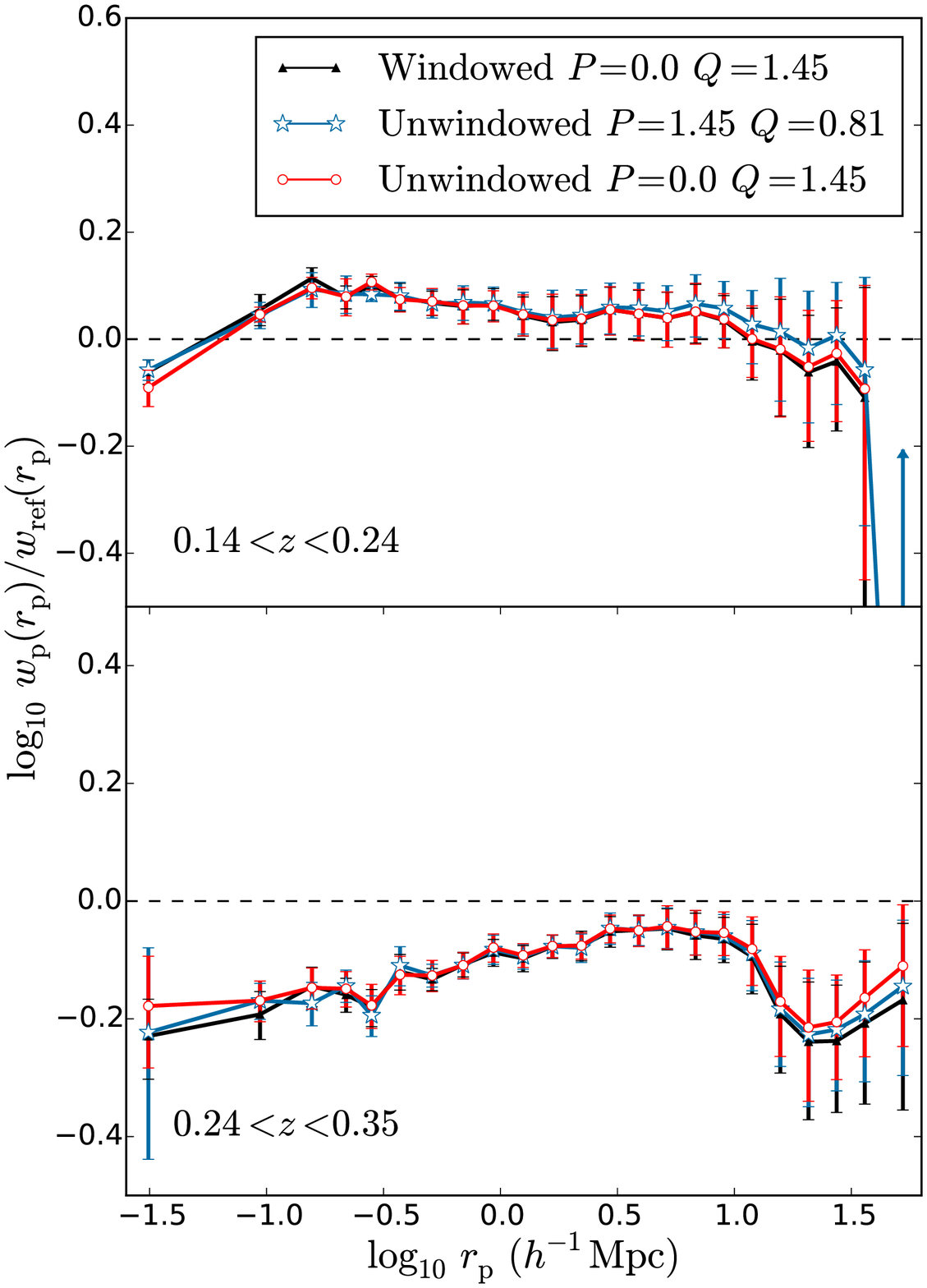}
\caption{ The projected 2PCF of $10.00<\log_{10}M_{*}/\mathrm{\mdot}h^{-2}<10.50$ galaxies in two different redshift slices, as indicated.
The different lines represent different methods of generating a random catalogue with which to measure clustering, each
is labelled in the legend. It can be seen that the our choice of random catalogue does not have a significant effect on 
our measurements.}\label{fig:cTestFig}
\end{figure}
In this paper we utilize a random catalogue generated using a new, windowed approach. We also use
a different combination of the luminosity function evolution parameters $Q=1.45, P=0.0$ to a different
set of parameters taken from an earlier draft of \citet{loveday15}: $Q=0.81, P=1.45$. To test for any biases introduced by this,
we compare results of using three different random catalogues: the randoms used in our measurements,
the same $P$ and $Q$ as our random catalogue but without the window function and a different $P$ and
$Q$ with and without the window function. For this purpose we use a mass selected 
sample, as this avoids changes in the sample definition arising due to a different value of $Q$. We choose the
$10.00<\log_{10}M_{*}/\mathrm{\mdot}h^{-2}<10.50$ mass bin, and use the redshift ranges $0.14<z<0.24$ and $0.24<z<0.35$. These
redshift slices encompass a large under-density in the $n(z)$, where one might expect to see the largest
differences between the random catalogues.

In Fig.~\ref{fig:cTestFig} we show the results. The most difference appears at large-scales; with the measurements
using the windowed catalogue showing a lower clustering amplitude than the results from the two random catalogues without windowing. The
two unwindowed results also show some deviation on large-scales. Note, however, that the differences are much smaller than the error bars 
for both of the samples, so much so that the conclusions of this paper will not be affected.  

\section{Power-law fit tables and mock sample properties}\label{sec:tables}
In Table~\ref{table:fits} we show the results from our power law fits to our samples of GAMA galaxies. Uncertainties are given in brackets, and were
derived from fitting power laws to each jack-knife realisation separately. Also given are the $r_{0}$ values found from fixing the 
slope of the power law, $\gamma$, rather than allow it to vary freely. The difference in the recovered $r_{0}$ values is small.

Tables~\ref{table:g14Samp} and \ref{table:l14Samp} give the sample properties, and standard deviations (in brackets) of one realisation of the
G14 and L14 models respectively. The magnitudes have been adjusted in order to match their luminosity functions to the real GAMA data 
(see~\ref{sec:modelcuts}), as such the sample sizes are much more similar to the real data for luminosity-defined samples than mass- or colour-defined 
samples.

\begin{table*}
\caption{The recovered values from power law fits to the clustering, $r_{0, \gamma\,\rm{fixed}}$ are the results of fitting the data with a 
fixed value of $\gamma=1.8$. Jack-knife errors are given in brackets. We see red galaxies have steeper 2PCFs, and that the amplitude of the 2PCF increases with magnitude 
and mass, regardless as to whether $\gamma$ is also fit or not. Samples marked with asterisks are magnitude limited, so should be treated with careful consideration
of the GAMA selection function. Samples without asterisks are volume limited, and can be treated at face value.}\label{table:fits}
\begin{tabular}{c c c c c c}
Sample & $z_{\mathrm{min}}$  & $z_{\mathrm{max}}$ & $r_{0}$ & $\gamma$ & $r_{0,\gamma\,\mathrm{fixed}}$ \\ 
\hline
\hline
 $-18.00<M_{r,h}<-17.00$ & 0.02 & 0.14* & 3.68 (0.31) & 1.81 (0.09) & 3.69 (0.27)\\
 \hline
 $-19.00<M_{r,h}<-18.00$ & 0.02 & 0.14* & 4.57 (0.37) & 1.74 (0.06) & 4.44 (0.26)\\
 $$ &  0.14 & 0.24* & 4.51 (0.69) & 1.84 (0.05) & 4.52 (0.70)\\
 \hline
 $-20.00<M_{r,h}<-19.00$ & 0.02 & 0.14 & 5.35 (0.51) & 1.70 (0.07) & 5.07 (0.36)\\
 $$ & 0.14 & 0.24* & 4.91 (0.22) & 1.78 (0.03) & 4.86 (0.17) \\
 $$ & 0.24 & 0.35* & 4.05 (0.17) & 1.69 (0.05) & 4.09 (0.18) \\
 \hline
 $-21.00<M_{r,h}<-20.00$ & 0.02 & 0.14 & 5.58 (0.75) & 1.76 (0.07) & 5.45 (0.53)\\
 $$ & 0.14 & 0.24 & 5.39 (0.28) & 1.78 (0.04) & 5.33 (0.19)\\
 $$ & 0.24 & 0.35* & 5.04 (0.09) & 1.76 (0.02) & 4.98 (0.07)\\
 $$ & 0.35 & 0.50* & 4.82 (0.27) & 1.62 (0.06) & 4.84 (0.26)\\
 \hline
 $-22.00<M_{r,h}<-21.00$ & 0.02 & 0.14 & 6.59 (0.58) & 1.77 (0.08) & 6.49 (0.44)\\
 $$ & 0.14 & 0.24 & 6.12 (0.32) & 1.87 (0.05) & 6.30 (0.25)\\
 $$ & 0.24 & 0.35 & 6.47 (0.18) & 1.79 (0.02) & 6.45 (0.16)\\
 $$ & 0.35 & 0.50* & 6.38 (0.25) & 1.81 (0.03) & 6.40 (0.20)\\
 \hline
 \hline
 $-23.00<M_{r,h}^{0.1}<-22.00$ & 0.01 & 0.50 & 9.39 (0.50) & 1.99 (0.10) & 9.56 (0.44)\\
 $-22.00<M_{r,h}^{0.1}<-21.00$ & 0.01 & 0.38 & 6.65 (0.10) & 1.77 (0.01) & 6.60 (0.09)\\
 $-21.00<M_{r,h}^{0.1}<-20.00$ & 0.01 & 0.26 & 5.72 (0.28) & 1.75 (0.03) & 5.55 (0.20)\\
 $-20.00<M_{r,h}^{0.1}<-19.00$ & 0.01 & 0.18 & 5.27 (0.42) & 1.74 (0.04) & 5.08 (0.31)\\
 $-19.00<M_{r,h}^{0.1}<-18.00$ & 0.01 & 0.12 & 3.97 (0.33) & 1.87 (0.06) & 4.09 (0.27)\\
 \hline
 \hline
 $8.50<\log_{10}M_{*}/\mathrm{\mdot}h^{-2}<9.50$ & 0.02 & 0.05 & 3.28 (0.42) & 2.24 (0.21) & 4.86 (0.38)\\ 
 $$ & 0.02 & 0.14* & 4.00 (0.22) & 1.73 (0.05) & 3.95 (0.21) \\
 $$ & 0.14 & 0.24* & 3.70 (0.19) & 1.66 (0.03) & 3.65 (0.17) \\
 \hline
 $9.50<\log_{10}M_{*}/\mathrm{\mdot}h^{-2}<10.00$ & 0.02 & 0.14 & 5.84 (0.64) & 1.77 (0.07) & 5.72 (0.37)\\
 $$ & 0.14 & 0.24* & 4.81 (0.24) & 1.77 (0.03) & 4.75 (0.20)\\
 $$ & 0.24 & 0.35* & 3.44 (0.11) & 1.64 (0.05) & 3.64 (0.09)\\
 \hline
 $10.00<\log_{10}M_{*}/\mathrm{\mdot}h^{-2}<10.50$ & 0.02 & 0.14 & 5.93 (0.71) & 1.82 (0.07) & 6.01 (0.54)\\
 $$ & 0.14 & 0.18  & 5.94 (0.46) & 1.86 (0.05) & 6.35 (0.36)\\
 $$ & 0.14 & 0.24* & 5.67 (0.27) & 1.85 (0.04) & 5.86 (0.17)\\
 $$ & 0.24 & 0.35* & 4.81 (0.13) & 1.73 (0.02) & 4.71 (0.11)\\
 $$ & 0.35 & 0.50* & 4.49 (0.20) & 1.69 (0.04) & 4.65 (0.19)\\
 \hline
 $10.50<\log_{10}M_{*}/\mathrm{\mdot}h^{-2}<11.00$ & 0.02 & 0.14 & 6.84 (0.62) & 1.75 (0.09) & 6.69 (0.47)\\
 $$ & 0.14 & 0.24 & 6.37 (0.34) & 1.93 (0.04) & 6.78 (0.30)\\
 $$ & 0.24 & 0.29 & 5.76 (0.28) & 1.93 (0.03) & 6.65 (0.24)\\
 $$ & 0.24 & 0.35* & 6.37 (0.20) & 1.84 (0.02) & 6.50 (0.17)\\
 $$ & 0.35 & 0.50* & 6.15 (0.22) & 1.78 (0.03) & 6.10 (0.16)\\
 \hline
 $11.00<\log_{10}M_{*}/\mathrm{\mdot}h^{-2}<11.50$ & 0.24 & 0.35 & 8.42 (0.37) & 1.76 (0.03) & 8.33 (0.34)\\
 $$ & 0.35 & 0.37  & 8.06 (1.33) & 2.14 (0.11) & 10.88 (1.19)\\
 $$ & 0.35 & 0.50* & 8.17 (0.47) & 1.88 (0.05) & 8.38 (0.44)\\
 \hline
 \hline
 $\rm{Red}$ & 0.02 & 0.14* & 6.25 (0.53) & 2.03 (0.07) & 7.13 (0.45)\\
 $$ & 0.14 & 0.24* & 6.37 (0.31) & 1.96 (0.03) & 7.04 (0.25)\\
 $$ & 0.24 & 0.35* & 6.49 (0.15) & 1.90 (0.01) & 6.79 (0.14)\\
 $$ & 0.35 & 0.50* & 7.50 (0.29) & 1.89 (0.02) & 7.82 (0.25)\\
 \hline
 $\rm{Blue}$ & 0.02 & 0.14* & 3.34 (0.14) & 1.52 (0.02) & 3.03 (0.10)\\
 $$ & 0.14 & 0.24* & 3.89 (0.14) & 1.63 (0.03) & 3.63 (0.10)\\
 $$ & 0.24 & 0.35* & 4.09 (0.11) & 1.64 (0.02) & 4.00 (0.10)\\
 $$ & 0.35 & 0.50* & 4.80 (0.10) & 1.69 (0.03) & 4.85 (0.10)\\
\end{tabular}
\end{table*}

\begin{table*}
\begin{center}
  \caption{Different mock galaxy samples, sample sizes and median properties for one G14 model realisation of GAMA. The subscript 
    `med' indicates the values are medians. Where maximum redshift has an asterisk, the sample is not volume limited. 
     In samples without an asterisk at least 95\% of the members are volume limited. Values in brackets are rms scatter.}
  \medskip
  \begin{tabular}{c c c c c c c c c}
    \hline\hline
     Sample  & $z_{\mathrm{min}}$  & $z_{\mathrm{max}}$ & $N_{\rm{gals}}$ &   $z_{\mathrm{med}}$  & $M_{\mathrm{med}}$ & $\log_{10}(M_{*}/\mathrm{\mdot}h^{-2})_{\mathrm{med}}$ & $(g-r)_{\mathrm{0, med}}$ \\
    \hline
    \hline
$-18.00<M_{r,h}<-17.00$ & 0.02 & 0.07 & 2965 &  0.05 & -17.46 (0.28) & 8.83 (0.26) & 0.29 (0.17)\\
$$ & 0.02 & 0.14* & 5907 &  0.07 & -17.59 (0.27) & 8.83 (0.25) & 0.27 (0.17)\\
\hline
$-19.00<M_{r,h}<-18.00$ & 0.02 & 0.11 & 7402 &  0.08 & -18.48 (0.29) & 9.20 (0.30) & 0.30 (0.18) \\
$$ & 0.02 & 0.14* & 12979 &  0.10 & -18.55 (0.27) & 9.21 (0.30) & 0.29 (0.18)\\
$$ & 0.14 & 0.24* & 2894 &  0.15 & -18.85 (0.12) & 9.28 (0.33) & 0.30 (0.19)\\
\hline
$-20.00<M_{r,h}<-19.00$ & 0.02 & 0.14 & 11689 &  0.11 & -19.47 (0.29) & 9.71 (0.37) & 0.43 (0.20)\\
$$ & 0.14 & 0.17 & 7029 &  0.15 & -19.45 (0.28) & 9.75 (0.39) & 0.51 (0.20) \\
$$ & 0.14 & 0.24* & 26387 &  0.19 & -19.61 (0.27) & 9.78 (0.39) & 0.49 (0.21)\\
$$ & 0.24 & 0.35* & 2343 &  0.25 & -19.91 (0.09) & 9.65 (0.29) & 0.27 (0.17)\\
\hline
$-21.00<M_{r,h}<-20.00$ & 0.02 & 0.14 & 6968 &  0.11 & -20.40 (0.28) & 10.26 (0.36) & 0.61 (0.21)\\
$$ & 0.14 & 0.24 & 22285 &  0.20 & -20.39 (0.28) & 10.28 (0.35) & 0.61 (0.21)\\
$$ & 0.24 & 0.35* & 35746 &  0.28 & -20.56 (0.26) & 10.24 (0.36) & 0.50 (0.21)\\
$$ & 0.35 & 0.50* & 3644 &  0.37 & -20.87 (0.14) & 10.05 (0.28) & 0.27 (0.17)\\
\hline
$-22.00<M_{r,h}<-21.00$ & 0.02 & 0.14 & 1835 &  0.11 & -21.28 (0.25) & 10.72 (0.32) & 0.70 (0.20)\\
$$ & 0.14 & 0.24 & 5838 &  0.20 & -21.28 (0.25) & 10.73 (0.32) & 0.70 (0.20)\\
$$ & 0.24 & 0.35 & 15354 &  0.30 & -21.29 (0.25) & 10.72 (0.32) & 0.66 (0.20)\\
$$ & 0.35 & 0.37 & 3316 &  0.36 & -21.29 (0.26) & 10.72 (0.32) & 0.64 (0.20)\\ 
$$ & 0.35 & 0.50* & 17073 &  0.40 & -21.41 (0.26) & 10.64 (0.35) & 0.48 (0.21)\\
\hline
\hline
$8.50<M_{*}<9.50$ & 0.02 & 0.05 & 3066 &  0.04 & -17.49 (0.89) & 8.92 (0.28) & 0.43 (0.16)\\
$$ & 0.02 & 0.14* & 20046 &  0.09 & -18.34 (0.76) & 9.11 (0.26) & 0.27 (0.14)\\
$$ & 0.14 & 0.24* & 7525 &  0.17 & -19.23 (0.30) & 9.37 (0.12) & 0.25 (0.07)\\
\hline
$9.50<M_{*}<10.00$ & 0.02 & 0.14* & 10713 &  0.11 & -19.44 (0.63) & 9.73 (0.14) & 0.54 (0.16)\\
$$ & 0.14 & 0.24* & 19616 &  0.19 & -19.79 (0.47) & 9.77 (0.14) & 0.35 (0.15)\\
$$ & 0.24 & 0.35* & 12697 &  0.28 & -20.34 (0.28) & 9.84 (0.12) & 0.27 (0.08)\\
\hline
$10.00<M_{*}<10.50$ & 0.02 & 0.14 & 5746 &  0.11 & -20.18 (0.58) & 10.20 (0.14) & 0.66 (0.15)\\
$$ & 0.14 & 0.18 & 5155 &  0.16 & -20.14 (0.57) & 10.20 (0.14) & 0.66 (0.15)\\
$$ & 0.14 & 0.24* & 18085 &  0.20 & -20.17 (0.53) & 10.22 (0.14) & 0.65 (0.15)\\
$$ & 0.24 & 0.35* & 19509 &  0.28 & -20.72 (0.42) & 10.25 (0.15) & 0.44 (0.16)\\
$$ & 0.35 & 0.50* & 8752 &  0.40 & -21.23 (0.30) & 10.24 (0.14) & 0.29 (0.09)\\
\hline
$10.50<M_{*}<11.00$ & 0.02 & 0.14 & 3455 &  0.11 & -20.71 (0.69) & 10.73 (0.14) & 0.76 (0.11)\\
$$ & 0.14 & 0.24* & 11227 &  0.20 & -20.69 (0.61) & 10.71 (0.14) & 0.75 (0.11)\\
$$ & 0.24 & 0.29* & 8747 &  0.27 & -20.83 (0.43) & 10.71 (0.14) & 0.72 (0.11)\\
$$ & 0.24 & 0.35* & 18649 &  0.29 & -20.94 (0.40) & 10.72 (0.14) & 0.71 (0.11)\\
$$ & 0.35 & 0.50* & 7885 &  0.39 & -21.39 (0.33) & 10.80 (0.15) & 0.67 (0.16)\\
\hline
$11.00<M_{*}<11.50$ & 0.24 & 0.35* & 3114 &  0.30 & -21.75 (0.50) & 11.09 (0.09) & 0.75 (0.05)\\
$$ & 0.35 & 0.37 & 609 &  0.36 & -21.79 (0.32) & 11.10 (0.09) & 0.74 (0.05)\\
$$ & 0.35 & 0.50* & 4377 &  0.41 & -21.84 (0.28) & 11.11 (0.09) & 0.73 (0.05)\\
\hline
\hline
$\rm{Red}$ & 0.02 & 0.14* & 19247 &  0.10 & -19.23 (1.37) & 9.87 (0.65) & 0.63 (0.10)\\
$(g-r)_{0} + 0.03(M_{r,h} - M_{r,h}^*)> 0.498$ & 0.14 & 0.24* & 31606 &  0.19 & -20.07 (0.70) & 10.35 (0.38) & 0.69 (0.09)\\
$$ & 0.24 & 0.35* & 28164 &  0.28 & -20.80 (0.51) & 10.67 (0.28) & 0.71 (0.08)\\
$$ & 0.35 & 0.50* & 10034 &  0.39 & -21.53 (0.38) & 10.97 (0.21) & 0.72 (0.06)\\
\hline
$\rm{Blue}$ & 0.02 & 0.14* & 24008 &  0.09 & -18.72 (1.27) & 9.18 (0.53) & 0.26 (0.07)\\
$(g-r)_{0} + 0.03(M_{r,h} - M_{r,h}^*)< 0.498$ & 0.14 & 0.24* & 26135 &  0.19 & -19.88 (0.70) & 9.68 (0.34) & 0.30 (0.08)\\
$$ & 0.24 & 0.35* & 26245 &  0.29 & -20.63 (0.49) & 10.01 (0.28) & 0.30 (0.09)\\
$$ & 0.35 & 0.50* & 12622 &  0.40 & -21.25 (0.40) & 10.26 (0.26) & 0.29 (0.10)\\
\end{tabular}\label{table:g14Samp}
\end{center}
\end{table*}

\begin{table*}
\begin{center}
  \caption{Different mock galaxy samples, sample sizes and median properties for one L14 model realisation of GAMA. The subscript 
   `med' indicates the values are medians. Where maximum redshift has an asterisk, the sample is not volume limited. 
     In samples without an asterisk at least 95\% of the members are volume limited.  Values in brackets are rms scatter.}
  \medskip
  \begin{tabular}{c c c c c c c c c}
    \hline\hline
    Sample  & $z_{\mathrm{min}}$  & $z_{\mathrm{max}}$ & $N_{\rm{gals}}$ &   $z_{\mathrm{med}}$  & $M_{\mathrm{med}}$ & $\log_{10}(M_{*}/\mathrm{\mdot}h^{-2})_{\mathrm{med}}$ & $(g-r)_{\mathrm{0, med}}$ \\
    \hline
    \hline
    $-18.00<M_{r,h}<-17.00$ & 0.02 & 0.07 & 2982 &  0.05 & -17.46 (0.28) & 8.82 (0.30) & 0.41 (0.15) \\
    $$ & 0.02 & 0.14* & 5922 &  0.07 & -17.59 (0.27) & 8.82 (0.29) & 0.39 (0.15)\\
    \hline
    $-19.00<M_{r,h}<-18.00$ & 0.02 & 0.11 & 7348 &  0.08 & -18.49 (0.29) & 9.21 (0.31) & 0.42 (0.15) \\ 
    $$ & 0.02 & 0.14* & 12815 &  0.10 & -18.55 (0.27) & 9.22 (0.31) & 0.41 (0.15)\\
    $$ & 0.14 & 0.24* & 2887 &  0.15 & -18.85 (0.12) & 9.27 (0.30) & 0.41 (0.16)\\
    \hline
    $-20.00<M_{r,h}<-19.00$ & 0.02 & 0.14 & 11665 &  0.11 & -19.47 (0.29) & 9.65 (0.35) & 0.51 (0.16)\\
    $$ & 0.14 & 0.17 & 7242 &  0.15 & -19.45 (0.29) & 9.67 (0.37) & 0.53 (0.17) \\
    $$ & 0.14 & 0.24* & 26443 &  0.19 & -19.61 (0.27) & 9.70 (0.38) & 0.53 (0.17)\\
    $$ & 0.24 & 0.35* & 2560 &  0.25 & -19.91 (0.09) & 9.62 (0.26) & 0.36 (0.15)\\
    \hline
    $-21.00<M_{r,h}<-20.00$ & 0.02 & 0.14 & 6934 &  0.11 & -20.41 (0.28) & 10.18 (0.40) & 0.62 (0.18)\\
    $$ & 0.14 & 0.24 & 22310 &  0.20 & -20.40 (0.28) & 10.22 (0.40) & 0.63 (0.18)\\
    $$ & 0.24 & 0.35* & 35739 &  0.28 & -20.56 (0.27) & 10.16 (0.41) & 0.56 (0.19)\\
    $$ & 0.35 & 0.50* & 3563 &  0.37 & -20.86 (0.15) & 10.00 (0.31) & 0.37 (0.16)\\
    \hline
    $-22.00<M_{r,h}<-21.00$ & 0.02 & 0.14 & 1833 &  0.11 & -21.28 (0.25) & 10.63 (0.39) & 0.65 (0.18)\\
    $$ & 0.14 & 0.24 & 5843 &  0.20 & -21.28 (0.25) & 10.65 (0.38) & 0.64 (0.18)\\
    $$ & 0.24 & 0.35 & 15349 &  0.30 & -21.28 (0.25) & 10.63 (0.38) & 0.61 (0.18)\\
    $$ & 0.35 & 0.37 & 3399 &  0.36 & -21.29 (0.26) & 10.64 (0.38) & 0.59 (0.17)\\
    $$ & 0.35 & 0.50* & 17320 &  0.40 & -21.42 (0.26) & 10.47 (0.39) & 0.49 (0.18)\\
    \hline
    \hline
    $8.50<M_{*}<9.50$ & 0.02 & 0.05* & 3077 &  0.04 & -17.46 (0.94) & 8.92 (0.28) & 0.55 (0.14)\\
    $$ & 0.02 & 0.14* & 19994 &  0.09 & -18.36 (0.80) & 9.12 (0.26) & 0.38 (0.12)\\
    $$ & 0.14 & 0.24* & 9080 &  0.17 & -19.29 (0.31) & 9.36 (0.12) & 0.36 (0.08)\\
    \hline
    $9.50<M_{*}<10.00$ & 0.02 & 0.14* & 11290 &  0.11 & -19.45 (0.71) & 9.72 (0.14) & 0.58 (0.13)\\
    $$ & 0.14 & 0.24* & 20525 &  0.19 & -19.86 (0.50) & 9.76 (0.14) & 0.44 (0.12)\\
    $$ & 0.24 & 0.35* & 15610 &  0.28 & -20.39 (0.31) & 9.82 (0.12) & 0.38 (0.09)\\
    \hline
    $10.00<M_{*}<10.50$ & 0.02 & 0.14 & 5550 &  0.11 & -20.09 (0.66) & 10.19 (0.14) & 0.68 (0.11)\\
    $$ & 0.14 & 0.18 & 5102 &  0.16 & -20.06 (0.65) & 10.20 (0.14) & 0.68 (0.12) \\
    $$ & 0.14 & 0.24* & 16659 &  0.20 & -20.14 (0.61) & 10.21 (0.14) & 0.67 (0.12)\\
    $$ & 0.24 & 0.35* & 18009 &  0.29 & -20.82 (0.47) & 10.22 (0.14) & 0.53 (0.12)\\
    $$ & 0.35 & 0.50* & 9983 &  0.40 & -21.33 (0.33) & 10.22 (0.14) & 0.42 (0.09)\\
    \hline
    $10.50<M_{*}<11.00$ & 0.02 & 0.14 & 2806 &  0.11 & -20.71 (0.58) & 10.71 (0.14) & 0.78 (0.10)\\
    $$ & 0.14 & 0.24* & 9615 &  0.20 & -20.66 (0.53) & 10.72 (0.14) & 0.78 (0.10)\\
    $$ & 0.24 & 0.29* & 7770 &  0.27 & -20.76 (0.44) & 10.72 (0.14) & 0.76 (0.10) \\
    $$ & 0.24 & 0.35* & 16067 &  0.29 & -20.88 (0.42) & 10.75 (0.14) & 0.75 (0.10)\\
    $$ & 0.35 & 0.50* & 6316 &  0.39 & -21.47 (0.38) & 10.78 (0.15) & 0.62 (0.14)\\
    \hline
    $11.00<M_{*}<11.50$ & 0.24 & 0.35* & 4280 &  0.30 & -21.42 (0.51) & 11.09 (0.09) & 0.79 (0.07)\\
    $$ & 0.35 & 0.37 & 760 &  0.36 & -21.50 (0.35) & 11.10 (0.09) & 0.78 (0.06) \\
    $$ & 0.35 & 0.50* & 3912 &  0.40 & -21.68 (0.35) & 11.12 (0.10) & 0.77 (0.08)\\
    \hline
    \hline
    $\rm{Red}$ & 0.02 & 0.14* & 20986 &  0.10 & -19.07 (1.38) & 9.80 (0.65) & 0.65 (0.09)\\
    $(g-r)_{0} + 0.03(M_{r,h} - M_{r,h}^*)> 0.548$ & 0.14 & 0.24* & 31557 &  0.19 & -20.06 (0.67) & 10.29 (0.41) & 0.70 (0.09)\\
    $$ & 0.24 & 0.35* & 27728 &  0.28 & -20.77 (0.47) & 10.68 (0.32) & 0.73 (0.08)\\
    $$ & 0.35 & 0.50* & 7944 &  0.39 & -21.46 (0.36) & 10.98 (0.24) & 0.75 (0.07)\\
    \hline
    $\rm{Blue}$ & 0.02 & 0.14* & 22218 &  0.10 & -18.80 (1.30) & 9.18 (0.53) & 0.36 (0.07)\\
    $(g-r)_{0} + 0.03(M_{r,h} - M_{r,h}^*)< 0.548$ & 0.14 & 0.24* & 26243 &  0.19 & -19.88 (0.73) & 9.62 (0.33) & 0.39 (0.08)\\
    $$ & 0.24 & 0.35* & 26870 &  0.29 & -20.65 (0.54) & 9.95 (0.29) & 0.40 (0.09)\\
    $$ & 0.35 & 0.50* & 14824 &  0.40 & -21.34 (0.43) & 10.22 (0.29) & 0.41 (0.10)\\

\end{tabular}\label{table:l14Samp}
\end{center}
\end{table*}

\end{document}